%% file: Krogager_Q2225_arXiv.tex
\documentclass[useAMS,usenatbib]{mn2e}
\usepackage{epsfig,natbib}
\usepackage{amssymb, amsmath}
\usepackage{graphicx, xcolor}
\usepackage{placeins, float}
\usepackage{times}

\newcommand\ion[2]{#1\,{\sc #2}}%  ion, i.e. CII = \ion{C}{ii}
\newcommand{\Av}{A(V)}
\newcommand{\Avabs}{${\rm A(V)_{DLA}}$}

\newcommand{\kms}{km~s$^{-1}$}

\usepackage{booktabs}

\title[A quasar reddened by a dusty DLA] {A quasar reddened by a sub-parsec sized, metal-rich and dusty cloud in a damped Lyman-$\bmath{\alpha}$ absorber at $\bmath{z=2.13}$
\thanks{Based on observations made with the Nordic Optical Telescope, operated by the Nordic Optical Telescope Scientific Association at the Observatorio del Roque
de los Muchachos, La Palma, Spain, of the Instituto de Astrof\'isica de Canarias. Based on Very Large Telescope observations carried out at the European Organisation for Astronomical Research in the Southern Hemisphere, Chile under DDT program 293.A-5033. Some of the data presented herein were obtained at the W.M. Keck Observatory, which is operated as a scientific partnership among the California Institute of Technology, the University of California and the National Aeronautics and Space Administration. The Observatory was made possible by the generous financial support of the W.M. Keck 
Foundation.}}

\author[Krogager et al.]{
J.-K. Krogager$^{1,2}$\thanks{E-mail:
krogager@dark-cosmology.dk},
J. P. U. Fynbo$^{1}$,
P. Noterdaeme$^{3}$,
T. Zafar$^{4}$,
P. M\o ller$^{4}$,\newauthor
C. Ledoux$^{2}$,
T. Kr\"uhler$^{2}$,
A. Stockton$^{5}$\\
$^{1}$Dark Cosmology Centre, Niels Bohr Institute, Copenhagen University,
Juliane Maries Vej 30, 2100 Copenhagen \O, Denmark\\
$^{2}$European Southern Observatory, Alonso de C\'ordova 3107, Vitacura, Casilla 19001, Santiago 19, Chile\\ 
$^{3}$Institut d'Astrophysique de Paris, CNRS-UPMC, UMR7095, 98bis bd Arago, 75014 Paris, France\\
$^{4}$European Southern Observatory, Karl-Schwarzschild-Stra{\ss}e 2,
85748 Garching bei M\"unchen, Germany\\
$^{5}$Institute for Astronomy, University of Hawaii, 2680 Woodlawn Drive, Honolulu, Hawaii, USA
}

\begin{document}

\date{Accepted Oct 6th, 2015.}

\pagerange{\pageref{firstpage}--\pageref{lastpage}} \pubyear{2014}

\maketitle

\label{firstpage}

\begin{abstract}

We present a detailed analysis of a red quasar at $z=2.32$ with an intervening damped Lyman-$\alpha$ absorber (DLA) at $z=2.13$. Using high quality data from the X-shooter spectrograph at ESO Very Large Telescope we find that the absorber has a metallicity consistent with Solar. We observe strong \ion{C}{i} and H$_2$ absorption indicating a cold, dense absorbing medium. Partial coverage effects are observed in the \ion{C}{i} lines, from which we infer a covering fraction of $27\pm6$~\% and a physical diameter of the cloud of 0.1~pc. From the covering fraction and size, we estimate the size of the background quasar's broad line region.
We search for emission from the DLA counterpart in optical and near-infrared imaging. No emission is observed in the optical data. However, we see tentative evidence for a counterpart in the $H$ and $K'$ band images.
The DLA shows high depletion (as probed by $[{\rm Fe/Zn}]=-1.22$) indicating that significant amounts of dust must be present in the DLA. By fitting the spectrum with various dust reddened quasar templates we find a best-fitting amount of dust in the DLA of ${\rm A(V)_{DLA}}=0.28\pm0.01|_{\rm stat}\ \pm0.07|_{\rm sys}$.
We conclude that dust in the DLA is causing the colours of this intrinsically very luminous background quasar to appear much redder than average quasars, thereby not fulfilling the criteria for quasar identification in the Sloan Digital Sky Survey. Such chemically enriched and dusty absorbers are thus underrepresented in current samples of DLAs.

\end{abstract}


\section{Introduction}

Damped Lyman-$\alpha$ absorbers (DLAs) are a class of neutral hydrogen absorbers seen in the spectra towards bright background sources (typically quasars and gamma ray bursts). Due to their high degree of self-shielding, these absorbers allow a precise characterization of the chemical evolution in the gas phase of galaxies over a large cosmic timespan \citep[for a review, see][]{Wolfe2005}. DLAs thereby play a significant role in understanding the interstellar medium (ISM) and star formation at high redshift.

However, in order to interpret the available data it is important to understand to which extent our quasar samples are biased against dusty and hence likely more metal-rich sightlines \citep{Pei1999}. This issue has been discussed extensively in the literature for more than 20 years \citep[e.g.][]{Pei1991, Vladilo2005, Pontzen2009, Khare2012}.
Over the past few years a handful of systems have been discovered in which the DLAs host significant amounts of dust \citep[e.g.][]{Fynbo2011, Noterdaeme2012a, Wang2012}. This poses a problem to the way we sample the population of absorbing galaxies.
As DLAs are detected towards bright background sources, the presence of dust in the absorbing medium will cause the background source to appear fainter and redder. Up until now, large area quasar samples have been compiled from optical photometry only (e.g., the Sloan Digital Sky Survey, SDSS; York et al. 2000)\nocite{York2000}. Hence optical colour criteria were the basis for selection algorithms of quasars. Recently the search for quasars has been refined and an increasing number of selection techniques are used to probe the quasar population to gather a more complete census of the quasars, and thereby also the absorbers along their sightlines \citep[e.g.,][]{Ross2012, Schmidt2010, Graham2014, Heintz2015}.
The importance of the dust bias can be studied by compiling a radio selected sample of quasars, since the radio emission is not affected by dust.
This approach has been followed both in the CORALS survey \citep{Ellison2001a, Ellison2004} and the UCSD survey \citep{Jorgenson2006}, which includes the CORALS survey. The authors of these surveys conclude that a dust bias is probably a minor effect, however, a larger sample is necessary to robustly exclude a significant dust bias.
The later study of \citet{Pontzen2009} confirms that a dust bias is most likely a small effect. Nevertheless, they find that the cosmic density of metals measured in DLA surveys could be underestimated by up to a factor of 2 due to a dust bias.

The effects from dust obscuration in a foreground absorber were identified clearly towards the quasar Q\,0918+1636 presented in the work by \citet{Fynbo2011}. The reddening of the background quasar, caused by dust in the foreground DLA, changed the colours of the quasar to the point where the quasar was close to dropping out of the SDSS quasar sample. In other words, had the dust extinction in the DLA been a little higher than the modest ${\rm A(V)}=0.2$~mag, the background quasar would not have been identified as a quasar in SDSS.
Following this discovery, we set out to look for quasars that would be missing in the SDSS quasar catalogs due to such reddening effects. The resulting surveys from our targeted search have identified almost 150 quasars that were not classified as such in SDSS-II \citep{Fynbo2013a, Krogager2015}. Although the vast majority of the quasars in these surveys were reddened by dust in the hosts rather than in foreground DLAs, we did encounter one very strong candidate: a DLA at redshift $z=2.13$ towards the quasar J\,2225+0527\footnote{Interestingly, this target, although not classified as such in SDSS, is a well-known radio source discovered by \citet*{Gower1967} and identified as 4C\,05.84. The QSO was later spectroscopically observed and a redshift of $z=2.323$ was measured \citep*{Barthel1990}. Barthel et al. also identified the $z_{\rm abs} = 2.132$ DLA. However, this did not come to our attention before starting the follow-up observations. The previous low resolution data were also not adequate for the detailed absorption line analysis in this work.}.

In this work, this specific target was then observed in greater detail at the Very Large Telescope to characterize the absorber and the extinction plausibly caused by it. We have combined the deep spectrum with new optical imaging from the Nordic Optical Telescope together with archival near-infrared imaging from the Keck observatory to search for the emission counterpart of the DLA. Such DLA galaxies have been exceedingly difficult to observe and the current sample of detections at redshift $z\approx2$ consists of only 10 galaxies \citep{Krogager2012}. It has been argued that the more evolved and metal-rich DLAs might also be associated with more massive systems and hence would be more readily detected in emission \citep{Moller2004,Fynbo2008,Neeleman2013}. In that case, metal-rich and dusty DLAs should be caused by the most evolved and massive systems, which are more likely to be missed in optically selected quasar samples.\\

This paper is structured as follows: In Section~\ref{data}, we present the data for this sightline, in Sections~\ref{absorption}, \ref{dust}, and \ref{emission}, we present the analysis of absorption lines, dust extinction, and the search for an emission counterpart, respectively, and in Section~\ref{discussion}, we discuss the implications of our findings.

Throughout this paper we assume a standard $\Lambda$CDM cosmology
with $H_0=67.9\, \mathrm{km s}^{-1}\mathrm{Mpc}^{-1}$, $\Omega_{\Lambda}=0.69$ and
$\Omega_{\mathrm{M}} = 0.31$ \citep{Planck2014}.

\section{Observations and data reduction}
\label{data}

\subsection{Spectroscopy}
We observed the target as a candidate quasar from the High \Av\ QSO (HAQ) survey \citep{Krogager2015} on the night of August 27, 2014 at the Nordic Optical Telescope. The initial observations were carried out using the Andalucia Faint Object Spectrograph and Camera (ALFOSC) with a low resolution setup. The DLA was identified during the observing run and was followed up with higher resolution spectra covering the absorber the following nights.

After the identification in the HAQ survey we observed the target with the X-shooter spectrograph \citep{Vernet2011} at the VLT during October 30, 2014 and November 23, 24, 25, 2014 under director's discretionary time program ID 293.A-5033. The X-shooter covers the wavelength range from $3000$~{\AA} to $2.5~\mu$m simultaneously by splitting the light into three separate spectrographs, the so-called arms: UVB ($3000-5500$~\AA), VIS ($5500-10000$~\AA), and NIR ($10000-25000$~\AA). All observations were carried out with the same slit widths (see Table~\ref{observinglog}) to allow us to search for emission from the DLA galaxy counterpart at small projected distances from the quasar. However, since the atmospheric dispersion correction did not function properly we were forced to observe at parallactic slit angle to limit slit loss. For this reason only two different slit angles were obtained, one for the Oct 30 observations (${\rm PA}\sim -4$~deg east of north) and three almost identical slit angles for the November observations (${\rm PA}\sim 39$~deg east of north).
The data were reduced using the official X-shooter pipeline {\tt esorex} version 2.3 for stare mode, whereafter the four individual 2-dimensional spectra were combined and the final 1-dimensional spectrum was extracted (this was separately done for each arm: UVB, VIS, NIR). This spectrum was then corrected for Galactic extinction ($E(B-V)=0.132$) using the maps of \citet{Schlafly2011}. The effective resolving power was measured in the final spectrum to be $R=11000$ for VIS and $R=7000$ for NIR. As there are no telluric lines appropriate for determining the resolution in the UVB, we obtained the resolving power by interpolating between the tabulated values of resolving power for given slit widths. Given the seeing of $\sim$0\farcs 7 in $V$-band we inferred a resolving power of $R\approx6000$ in the UVB .

\subsection{Imaging}
In order to look for extended emission from the galaxy counterpart of the DLA, we observed the region around the quasar with the ALFOSC instrument at the NOT in imaging mode. We used the $r'$ filter centred at 618~nm. The observations were carried out in a five-point dither pattern with sub-integrations of 330~s over many nights during November 2014 and June 2015 under program P49-422 (the dates are given in Table~\ref{observinglog}). The images were bias subtracted and flat field corrected using standard {\sc iraf} routines. In order to subtract fringes and large scale variations in the background we subsequently created a master background image. In each frame, we masked out any sources above 4 $\sigma$ of the sky noise estimated by the robust median absolute deviation. The masked frames were normalized by the median of each frame and combined. The resulting background image was then scaled to and subtracted from each frame. For each frame, we measured the seeing and only the 50 per cent best frames (21 out of 42) in terms of seeing were aligned and median combined to create the final image. This resulted in a much sharper and more symmetric point spread function (PSF) compared to the image of all frames combined.
The seeing in the optimally combined image was 0\farcs 74 compared to the 0\farcs 85 of the combined image from all frames.

Furthermore, we have included imaging data in the $J$, $H$, and $K'$-bands from the Keck II telescope observed with the NIRC2 instrument fed by the Keck Laser-Guide-Star Adaptive-Optics system to enhance the image resolution. The observations in $H$ and $K'$ bands had total exposure times of 2700~s each, performed in sub-integrations of 300~s. The $J$ band image was obtained with sub-integrations of 180~s for a total exposure time of 1620~s.
The NIRC2 images were reduced using an {\sc iraf} scripted pipeline that does an iterative removal of the sky background, flat fielding, registration of the individual images, and drizzling onto the final grid with a pixel size of 40~mas~pixel$^{-1}$.
The full width at half maximum (FWHM) of the point spread function in the $H$ and $K'$ images are 0\farcs089 and 0\farcs092, respectively. 
The shallower $J$-band image was obtained with slightly worse AO correction resulting in a less symmetric point spread function with a FWHM of 0\farcs093. Moreover, the wings of the PSF were more dominant for the $J$-band image than in the $H$ and $K'$-band images.

\begin{table}
\small
\caption{Observations of the quasar J\,222514.97+052709.1\label{observinglog}}
\begin{center}
\begin{tabular}{@{}l c l @{}r@{}}
\toprule
Date    &    Instrument    &    Setup    & Exp. Time    \\
        &                  &             &    (s)       \\[0.4ex]
\midrule
2006, Sep 1     &   Keck/NIRC2 AO     &    $H$, $K'$ imaging        &   $2700$ \\
2007, Aug 21    &   Keck/NIRC2 AO     &    $J$ imaging              &   $1620$ \\
2014, Oct 30    &   VLT/X-shooter &    1\farcs 3, 1\farcs 2, 1\farcs 2 $^{(a)}$   & $3000$   \\
2014, Nov 23    &   VLT/X-shooter &    1\farcs 3, 1\farcs 2, 1\farcs 2 $^{(a)}$   & $3000$   \\
2014, Nov 24    &   VLT/X-shooter &    1\farcs 3, 1\farcs 2, 1\farcs 2 $^{(a)}$   & $3000$   \\
2014, Nov 25    &   VLT/X-shooter &    1\farcs 3, 1\farcs 2, 1\farcs 2 $^{(a)}$   & $3000$   \\
2014, Nov 13, 14, 15  &   NOT/ALFOSC  &   $r'$-band imaging    &   $8580$    \\
2015, Jun 12, 23 &   NOT/ALFOSC  &   $r'$-band imaging    &   $5280$    \\
\bottomrule
\end{tabular}
\end{center}
$^{(a)}$ The slit widths of the three arms of X-shooter given as UVB, VIS and NIR.
\end{table}

\section{Absorption Line Analysis}
\label{absorption}
\begin{figure}
  \includegraphics[width=0.48\textwidth]{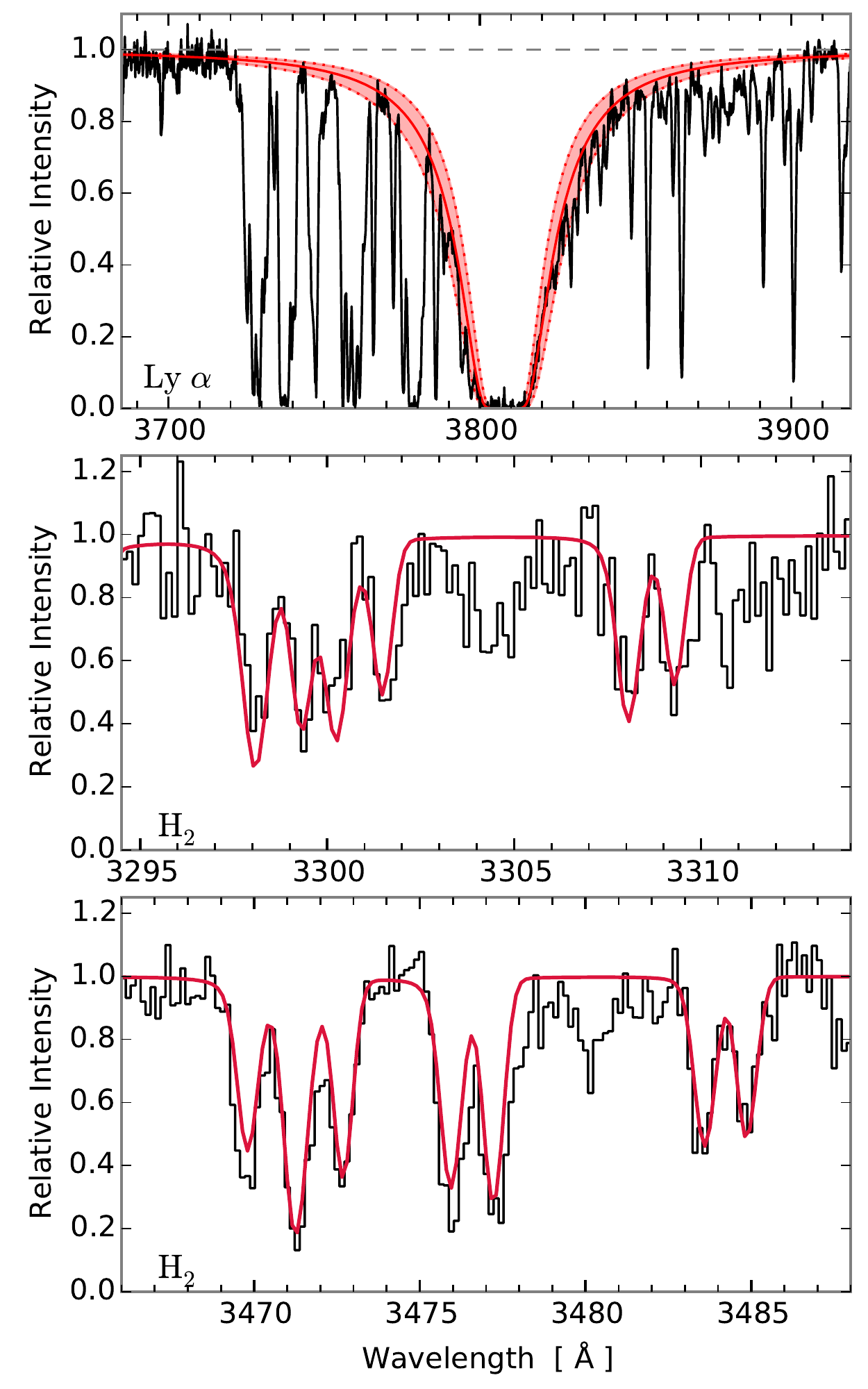}
  \caption{In the top panel, the Ly$\alpha$ transition is shown. The best-fitting Voigt profile is shown in solid red with 3-$\sigma$ range indicated by the red shaded area. In the middle and lower panels, a zoom-in around some of the available H$_2$ lines is shown. The red line marks the best-fitting model to the ro-vibrational lines.
   \label{fig:H_lines}}
\end{figure}

We detected molecular hydrogen in absorption from rovibrational transitions in the rest-frame far-UV. We fitted the Lyman and Werner bands of H$_2$ assuming two components for the $J=0,1,2,3$ levels. The relative velocities and broadening parameters were kept fixed for all $J$-levels. For the first component we inferred $z_1=2.13123$ ($v_{\rm rel}=-120$~km~s$^{-1}$) and $b=1.3$~km~s$^{-1}$, and for the second component we inferred $z_2=2.13240$ ($v_{\rm rel}=-9$~km~s$^{-1}$) and $b=8.1$~km~s$^{-1}$. Due to the poor resolution of the UVB arm, the very narrow molecular lines were not resolved and hence the fit was very degenerate. We therefore do not state all the column densities of each individual transition. However, the total column density remained rather unaffected of the degeneracies as it was dominated by only one component in the damped regime. From the fit we obtained a total column density of molecular hydrogen of $\log\,N({\rm H}_2)/{\rm cm}^{-2}=19.4\pm 0.1$. We observed tentative evidence of partial coverage (this effect is described in more detail in Section~\ref{partial_coverage}) from the H$_2$ lines at the Ly$\beta$ and \ion{O}{vi} lines, but higher resolution spectroscopy is needed to confirm.
In order to measure the column density of neutral hydrogen, we fitted the Ly$\alpha$ transition, for which we obtained $\log\,N({\rm H\,{\textsc i}})/{\rm cm}^{-2} = 20.69\pm0.05$. The molecular fraction $f_{{\rm H}_2}=2N({\rm H}_2)/(2N({\rm H}_2) + N({\rm H\, {\textsc i}})) = 0.09$ is thus among the highest observed in high-redshift quasar-DLAs \citep[e.g.][]{Noterdaeme2008, Srianand2008a, Noterdaeme2010, Noterdaeme2015b, Jorgenson2014}.
The fitted profiles to neutral and molecular hydrogen are shown in Fig.~\ref{fig:H_lines}.\\

We detected metal absorption lines from various ionization states, ranging from neutral species (\ion{C}{i}, \ion{Cl}{i}) to highly ionized transitions (\ion{C}{iv}, \ion{Si}{iv}). The full list of absorption lines detected and their equivalent widths are given in Table~\ref{full_linelist} of the Appendix. In order to obtain the column densities for the low ionization lines, we fitted the weak unsaturated lines using Voigt profiles. We used the iron and manganese lines (\ion{Fe}{ii}$\lambda2374$ and \ion{Mn}{ii}$\lambda2576$) with high signal-to-noise ratios to obtain a solution for the number of velocity components, their relative velocities, and their broadening parameters. The best-fitting solution to these lines gave 10 components ranging from $v=-183$ to $+214$~km~s$^{-1}$ relative to the strongest absorption component, $z_{\rm sys}=2.13249$. The relative velocities and broadening parameters of the individual components are given in Table~\ref{tab:fit_details} for reference. We assumed the broadening to be purely turbulent. This velocity structure was then kept fixed and all lines were fitted simultaneously to infer their column densities. We included the uncertainty on the continuum normalization in the fitting. The best-fitting column densities are reported in Table~\ref{tab:fit_details}. The neutral magnesium transition \ion{Mg}{i} $\lambda2852$ was included in order to constrain the contribution from Mg in the line-blend of Zn, Cr, and Mg at 2026~{\AA}. Since the signal-to-noise for the \ion{Mg}{i} line was high, we fitted a separate set of 7 components to this line. Because of the different velocity structure of the magnesium line we do not include the individual components in Table~\ref{tab:fit_details}. We further note that the contribution from \ion{Co}{ii} to the line-blend is negligible due to the weak strength of this transition and the non-detection of other Co transitions, see Table~\ref{full_linelist}. For lines with lower signal-to-noise ratios, we were not able to constrain all 10 components. These unconstrained components are marked with `\ldots' in Table~\ref{tab:fit_details}. The total abundances from the fit and the calculated metallicities are given in Table~\ref{LineList}. We observe a relatively low [S/Zn] ratio of $-$0.31, this is lower than the average DLA, however not uncommon given the large uncertainty regarding this element \citep{Rafelski2012, Berg2015}. Furthermore, the abundance of sulphur may be underestimated due to hidden saturation given the lower resolution in the UVB arm. For a resolution of $\mathcal{R}=6000$, we find that absorption lines with a flux residual at peak absorption of less than roughly 0.75 are potentially affected by hidden saturation. One of the \ion{S}{ii} lines is thus right at the border, but this line is blended with Ly$\alpha$ forest. The other sulphur line is stronger and should therefore be viewed as a lower limit.
The hidden saturation might cause the column density to be underestimated by as much as 0.4~dex.

From the low ionization metal lines, we were able to recover information about the kinematics of the DLA through the measure $\Delta V_{90}$. We followed the definition by \citet*{Prochaska1997}, who define $\Delta V_{90}$ in terms of the wavelengths encompassing the central 90 per cent of the apparent optical depth.
In order to avoid saturation effects biasing the apparent optical depth we chose the \ion{Si}{ii} $\lambda1808$ absorption line instead of the stronger \ion{Fe}{ii} $\lambda2374$ transition. Following the definition by Prochaska \& Wolfe, we first converted the normalized spectral line to apparent optical depth:
$\tau = -\ln\ (F)$, where $F$ denotes the continuum normalized flux.
From the distribution of $\tau$ we then calculated the line centre, $\lambda_0$ along with the 5th and 95th percentiles, $\lambda_5$ and $\lambda_{95}$, respectively (see Fig.~\ref{fig:vel_width}). To minimize the contribution from noise in the continuum, we only calculated the percentiles within a set of boundaries obtained from visual inspection of the absorption line profiles. The boundaries used for this analysis were $-250$ and $+290$~km~s$^{-1}$. We then obtained the velocity spread as:
$$\Delta V_{90} = c\, (\lambda_{95} - \lambda_{5})/\lambda_0 = 331\pm30~{\rm km~s}^{-1}~.$$
We note that the result does not depend on the extent of the boundaries; only the uncertainty increases for more inclusive boundaries. The uncertainty on $\Delta V_{90}$ was determined by varying the spectral profile within the errors 10000 times.
For each realization, we measured $\Delta V_{90}$. The 1 $\sigma$ uncertainty was then inferred from the distribution of velocities as the 16th and 84th percentiles.
Furthermore, we obtained consistent results for the \ion{Fe}{ii} line. However, the possible hidden saturation affecting absorption lines with a flux residual at peak absorption of less than 0.6 yields the \ion{Fe}{ii} less reliable (given the resolution of 27.3~km~s$^{-1}$).

\begin{figure}
  \includegraphics[width=0.48\textwidth]{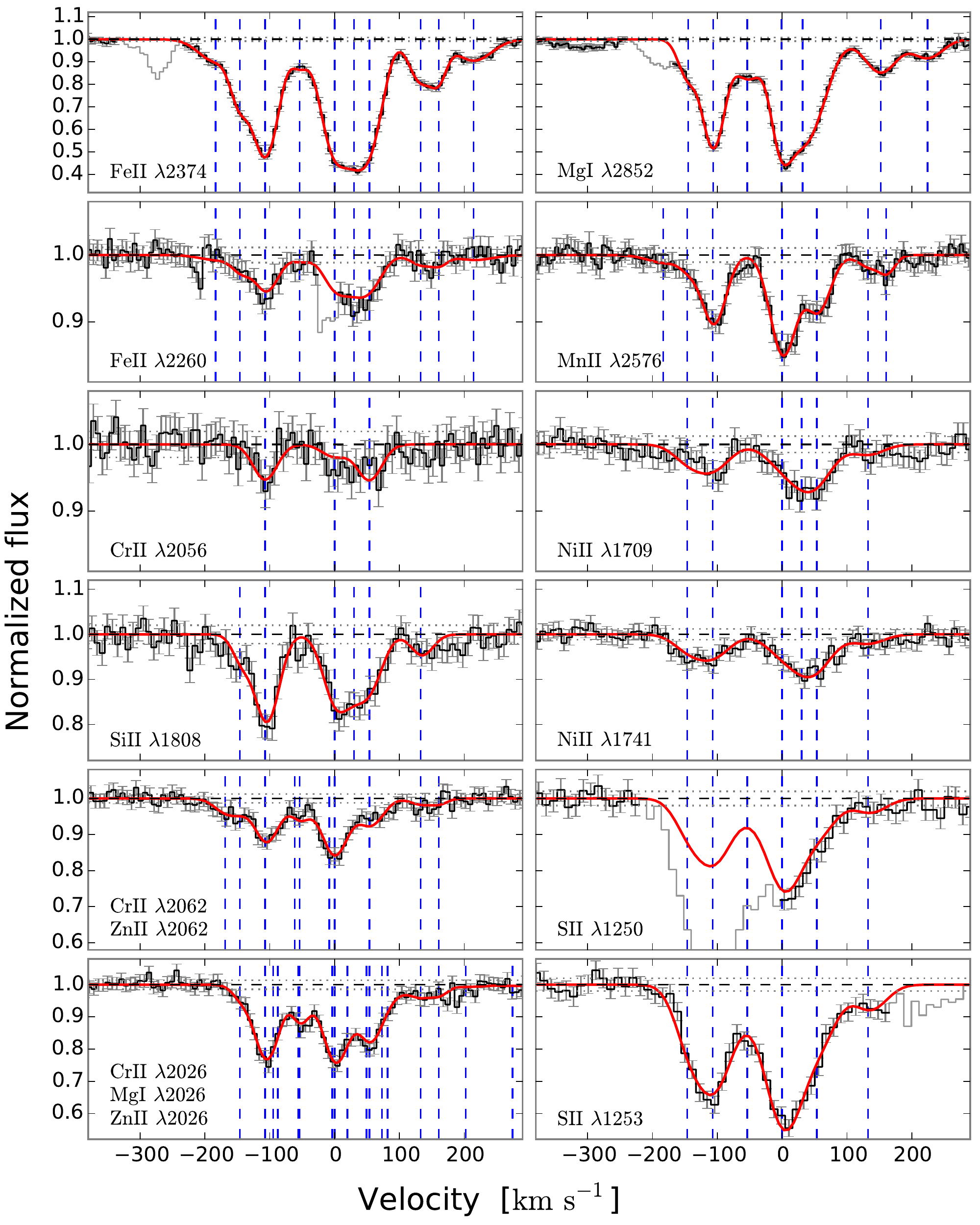}
  \caption{Voigt profile fits to low ionization lines. The data are shown in black with error-bars. Grey areas were excluded in the fits due to blends with other lines. The red line shows the best-fitting model, and the blue dashed lines show the fitted components. Velocities are relative to $z_{\rm sys}=2.13249$.
   \label{fig:abs_lines}}
\end{figure}

\begin{figure}
  \includegraphics[width=0.48\textwidth]{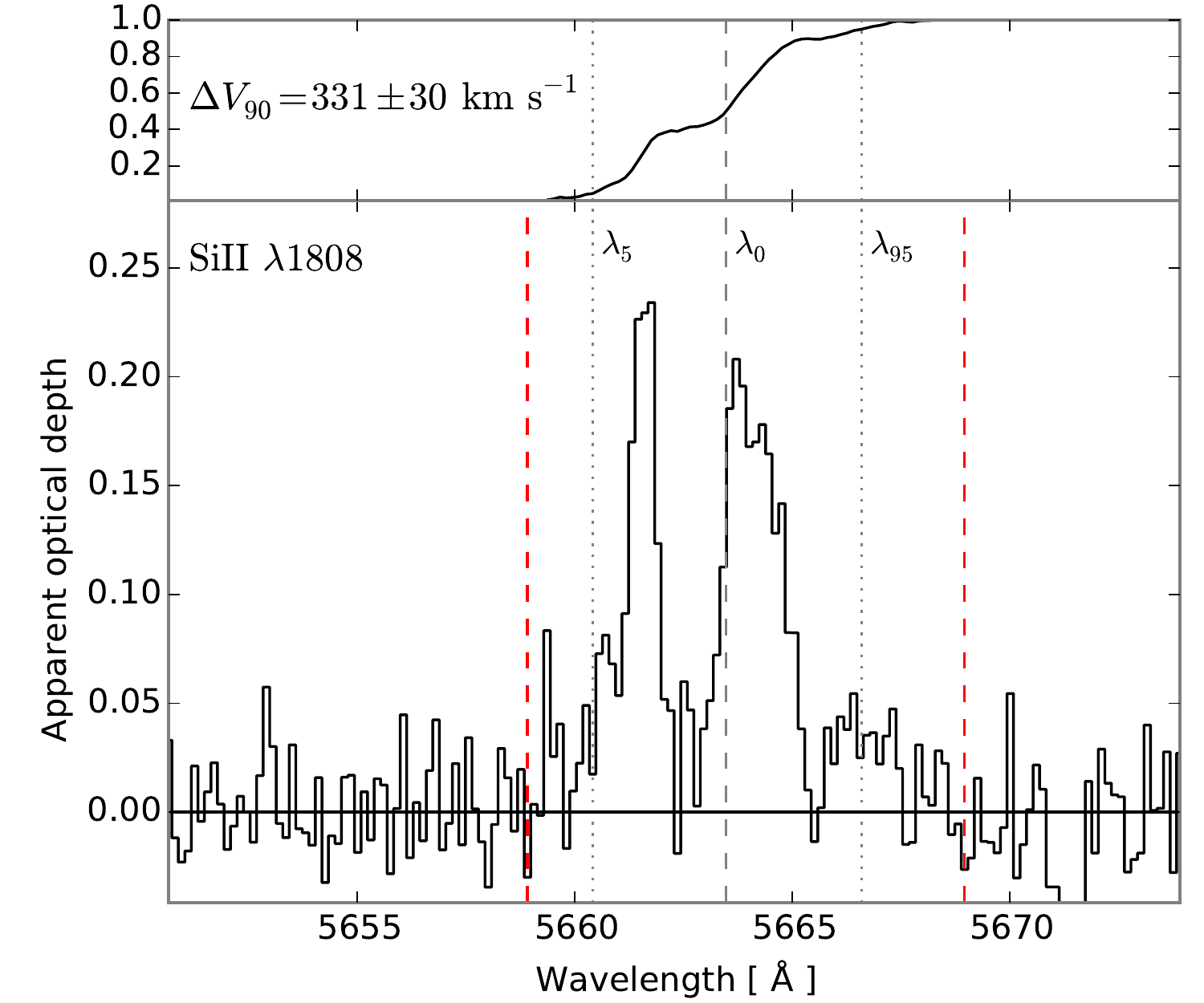}
  \caption{Apparent optical depth of the \ion{Si}{ii} $\lambda$1808 line used to calculate $\Delta V_{90}$. In the top panel, we show the cumulative distribution of apparent optical depth. The grey dashed line marks the absorption line centre, and the grey dotted lines mark the 5th and 95th percentiles of this distribution, from which $\Delta V_{90}$ is calculated. The red dashed lines mark the boundaries of the absorption line within which the apparent optical depth distribution is calculated. The boundaries used are $-250$ and $290$~km~s$^{-1}$ with respect to the systemic redshift, $z_{\rm sys}=2.13249$.
  \label{fig:vel_width}}
\end{figure}

%% TABLES
\input{metal_table}

\begin{table}
\caption{Metal abundances from the $z=2.13249$ DLA. \label{LineList}}
\begin{center}
\begin{tabular}{l c c}
\hline
\hline
Element ($X$) & $\log\left(N_X/{\rm cm}^{-2}\right)$ & [X/H]$^{(a)}$ \\
\hline
\ion{Si}{ii}, $\lambda1808$	            &  $15.49\pm0.04$  &   $-0.71\pm0.07$ \\
\ion{S}{ii}, $\lambda\lambda1250,1253$	&  $15.41\pm0.02$  &   $-0.40\pm0.06$ \\
\ion{Cr}{ii}, $\lambda\lambda\lambda2026,2056,2062$	& $13.04\pm0.08$ & $-1.29\pm0.09$ \\
\ion{Mn}{ii}, $\lambda2576$	            &  $12.91\pm0.02$  &   $-1.21\pm0.06$ \\
\ion{Fe}{ii}, $\lambda\lambda2374,2260$	&  $14.87\pm0.01$  &   $-1.32\pm0.05$ \\
\ion{Ni}{ii}, $\lambda\lambda1709,1741$	&  $13.90\pm0.07$  &   $-1.01\pm0.10$ \\
\ion{Zn}{ii}, $\lambda\lambda2026,2062$	&  $13.16\pm0.02$  &   $-0.09\pm0.05$ \\
             &               &    \\
\ion{C}{i}, $\lambda\lambda1560,1656$   &	$13.86\pm0.04$ & \ldots \\
\ion{C}{i}$^*$     	                    &	$13.80\pm0.04$ & \ldots \\
\ion{C}{i}$^{**}$        	            &	$12.98\pm0.11$ & \ldots \\
\ion{Cl}{i}, $\lambda1347$	            &   $13.34\pm0.03$ & \ldots \\
\ion{Mg}{i}, $\lambda\lambda2026,2852$	&   $12.93\pm0.02$ & \ldots \\
H$_2$	&   $19.4\pm0.1$ & \ldots \\

\hline
\end{tabular}
\end{center}

$^{(a)}$ With respect to solar abundances from \citet{Asplund2009}.\\

\end{table}

\subsection{Atomic Carbon}
\label{partial_coverage}
Strong absorption from atomic carbon (\ion{C}{i}) and its fine-structure levels $^3P_1$ and $^3P_2$ (here denoted as \ion{C}{i}$^*$ and \ion{C}{i}$^{**}$, respectively) was clearly detected in the rest-frame UV. We fitted these transitions with a two component Voigt profile for each of the three levels ($v_1=-110$~km~s$^{-1}$ and $v_2=-4$~km~s$^{-1}$, relative to $z_{\rm sys}=2.13249$). Due to the limited resolution of the spectra we fixed the relative velocities and doppler broadening parameters of the various transitions and fitted them all simultaneously. We were not able to get both the transitions at $1560$~\AA\ and $1656$~\AA\ to fit simultaneously. The depth of the absorption lines was over-estimated for the $1656$~\AA\ line while under-estimated for the $1560$~\AA\ line. This may be explained due to so-called partial coverage. The effect arises when only a fraction of the emitted flux from the background source passes through the absorbing medium, i.e., the absorbing medium has a covering fraction less than unity \citep[see][]{Balashev2011}.
The partial coverage leads to an observable effect where saturated absorption lines do not reach zero flux since a fraction of background emission is able to reach the observer unabsorbed. This excess flux is here referred to as the line flux residual, LFR. We follow Balashev et al (2011) in their definition of LFR. We call the total emitted flux $F_{\rm tot}(\lambda)$, and the flux that is affected by absorption $F_{\rm cloud}(\lambda)$. The observed flux is thus given by
$$F_{\rm obs}(\lambda) = F_{\rm tot}(\lambda) - F_{\rm cloud}(\lambda) \cdot [ 1 - e^{-\tau(\lambda)}] ~,$$
\noindent
where $\tau (\lambda)$ is the optical depth of the absorption line. The LFR is the normalized flux which is not covered by the cloud:
$$ {\rm LFR} = (F_{\rm tot} - F_{\rm cloud})/F_{\rm tot} ~.$$
\noindent
The phenomenon may also arise if the background source is gravitationally lensed and not all deflected paths pass through the absorbing medium; however, as will be argued later on, we disregard this effect in our analysis.

\begin{figure}
  \includegraphics[width=0.48\textwidth]{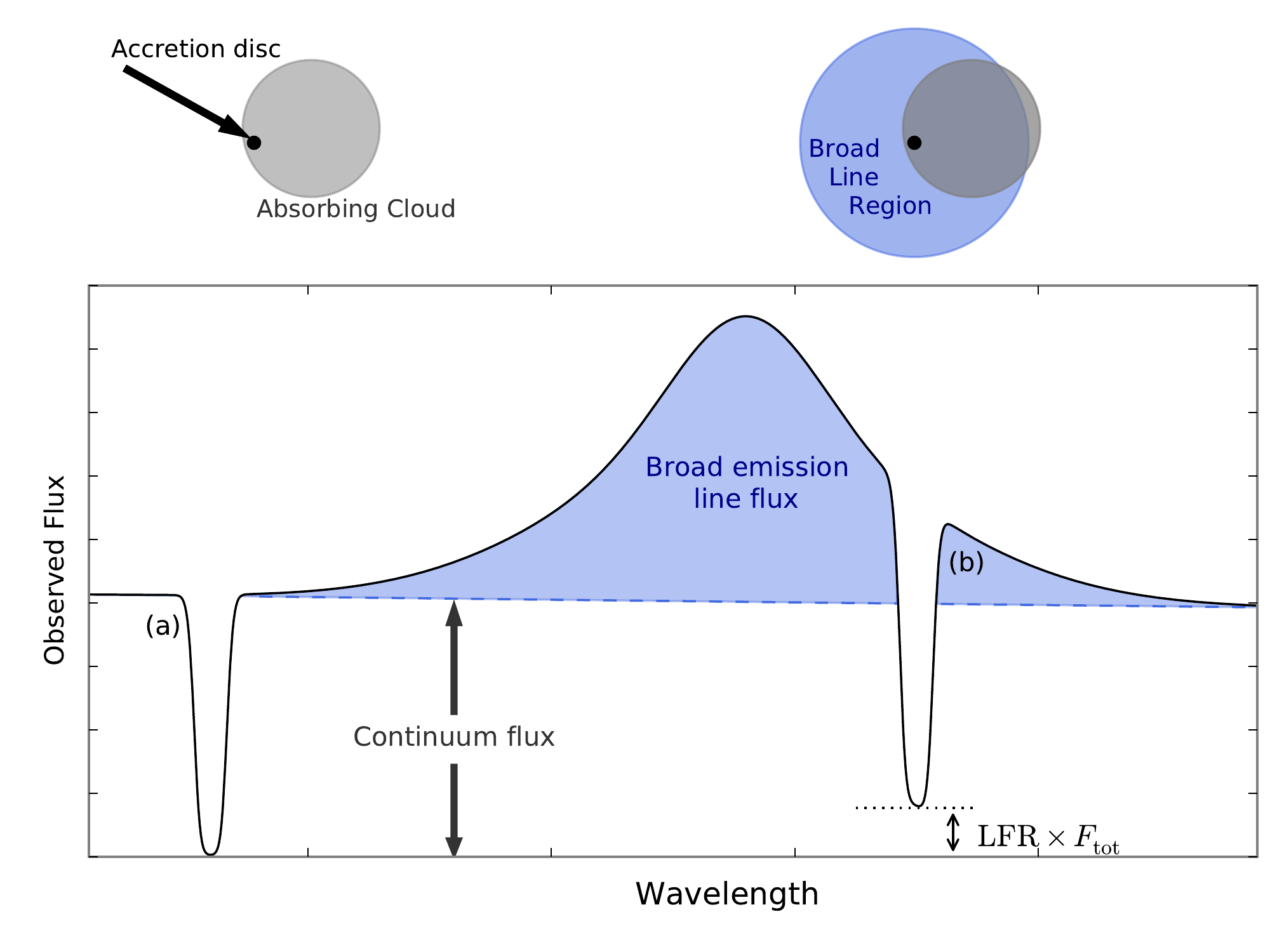}
  \caption{Schematic view of partially absorbing cloud. The top left panel shows the projected view of the background quasar's accretions disc (black dot) and the foreground absorbing cloud (grey circle). The top right panel shows the same configuration but with the quasar's broad line region shown as the blue circle. Absorption from different transitions in the foreground cloud are seen in the observed spectrum in the lower panel. Here the line marked (a) corresponds to the top left scenario at a wavelength where there is no contribution from broad emission line flux. The line marked (b) shows the top right scenario at a wavelength where broad emission line flux is contributing to the total observed flux. Due to the partial coverage of the broad line region indicated in the top right panel, a residual line flux (LFR) is observed in the saturated line, i.e., the line does not reach zero.
  \label{fig:CI_overview}}
\end{figure}

We assume a structure for the quasar as follows: the accretion disc around the black hole is responsible for the continuum emission and is assumed to be a point source; the broad emission lines are emitted from a more extended structure of gas called the broad line region (BLR). In this geometry, an absorption line on top of a continuum region will be unaffected as the cloud covers all of the continuum source. However, an absorption line on top of a broad emission line will absorb the full continuum source but only part of the BLR flux due to the partial coverage of the BLR. Absorption lines on top of broad emission lines will therefore exhibit a line flux residual depending on the covering fraction of the BLR. An idealised schematic view of the assumed geometry is shown in Fig~\ref{fig:CI_overview}. The figure shows the case of an absorption line on top of a broad emission line and one in a continuum region only. For clarity, we show saturated lines in the figure, as this allows the LFR to be visualized more easily. 

In case of gravitational lensing, the LFR for absorption lines in continuum regions and on top of broad emission lines will be identical, since the light from both the central continuum source and the BLR is deflected and reach the observer unabsorbed. Only in special cases, where the extended BLR is lensed differently from the continuum source, will the LFR be different. We therefore assume that the mismatch observed in the two \ion{C}{i} lines is caused by partial coverage due to the absorbing cloud being small in projection compared to the BLR.

\begin{figure}
  \includegraphics[width=0.48\textwidth]{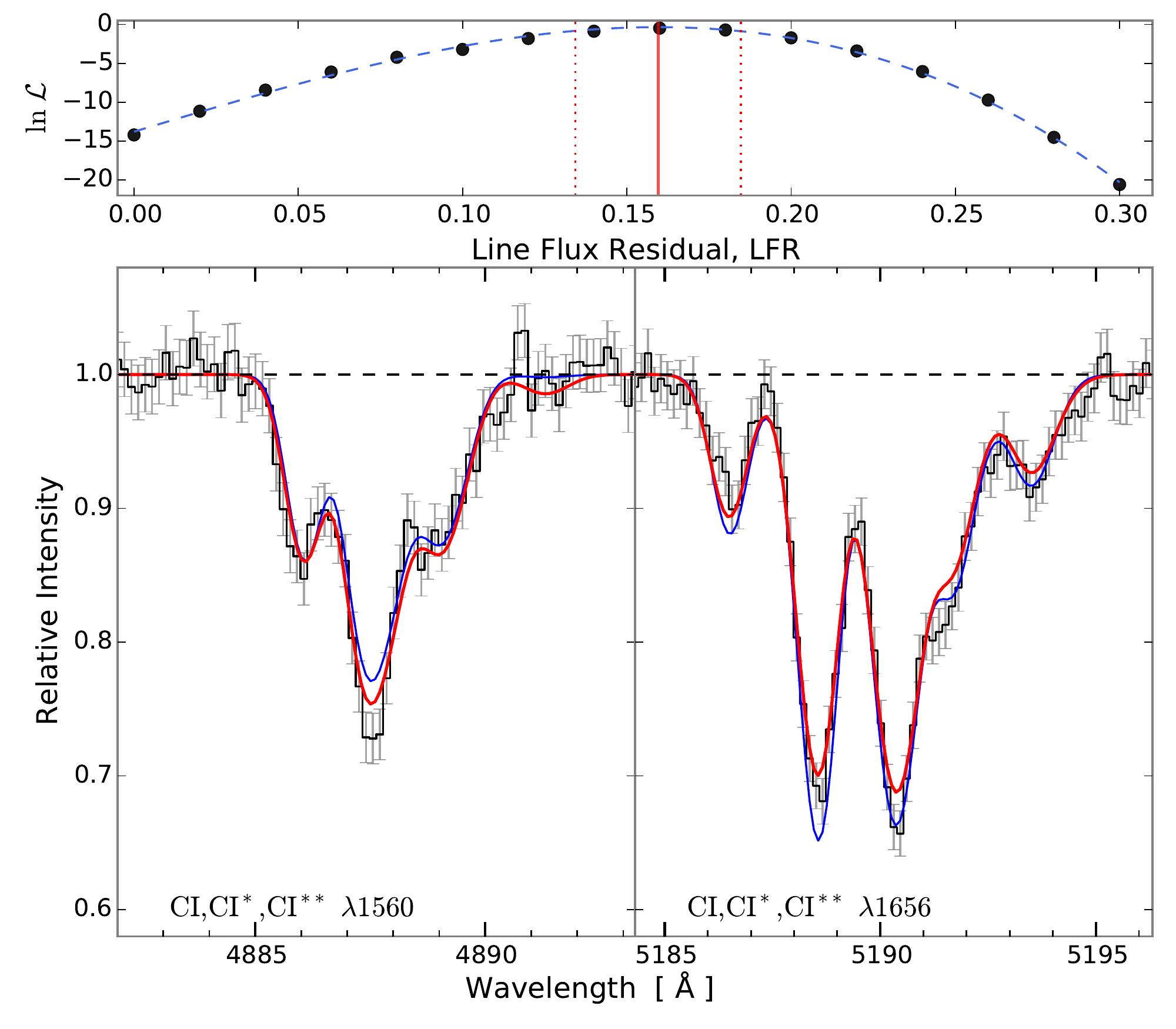}
  \caption{Voigt profile fit to neutral carbon (\ion{C}{i}) and its fine structure transitions (\ion{C}{i}$^*$ and \ion{C}{i}$^{**}$). The thick red line and thin blue line show the best-fitting model with and without partial coverage, respectively. The model without partial coverage poorly estimates the depth of the lines.
  The top panel shows the likelihood, $\mathcal{L}$ (normalized to the maximum likelihood) as function of the line flux residual, LFR, for each evaluation in black points. The blue dashed line shows a polynomial fit to the likelihood function. The maximum likelihood estimate (${\rm LFR}=0.16\pm0.03$) is shown by the red vertical line with 1~$\sigma$ confidence interval shown by the dotted lines.
  \label{fig:CI_fit}}
\end{figure}

Since the \ion{C}{i} lines were not saturated, we could not easily measure the LFR as is indicated in Fig~\ref{fig:CI_overview}. Instead we had to infer the LFR by fitting the absorption lines with various values for the LFR. We defined a grid of discrete value of LFR and for each grid point, we fitted the two absorption line complexes. The likelihood for each grid point was saved and used to fit a polynomial for the likelihood as function of LFR. We then obtained the best-fitting LFR by identifying the maximum likelihood from the polynomial fit, and the uncertainty was given by the curvature of the likelihood function, see top panel in Fig.~\ref{fig:CI_fit}. The lower panel in Fig.~\ref{fig:CI_fit} shows the best-fitting \ion{C}{i} absorption profile for the two transitions at 1560 and 1656~\AA\ in red. The blue line in the figure shows the fit without taking into account partial coverage. The best fit resulted in a line flux residual of $16\pm3$\,\%. Fitting the absorption lines for this value of the LFR, we recovered the column densities for \ion{C}{i}, \ion{C}{i}$^*$, and \ion{C}{i}$^{**}$ as given in Table~\ref{LineList}.

The $1560$~\AA\ transition is located in a part of the spectrum where the total incident flux is coming from the continuum source in the quasar ($F_{\rm cloud}(1560) = F_{\rm cont}$). The other transition ($1656$~\AA) is located on top of the broad \ion{C}{iv} emission line from the quasar; the incident flux is therefore a combination of the continuum flux and the flux from the more extended broad line region which is only partially covered ($F_{\rm cloud}(1656) = F_{\rm cont} + f_{\rm cov} \cdot F_{\rm BLR}$). The $F_{\rm cov}$ here denotes the fraction of the BLR covered by the cloud.
For the assumed quasar geometry, we can then relate $f_{\rm cov}$ to the estimated LFR:

\begin{equation}
\begin{split}
{\rm LFR} \cdot F_{\rm tot} & = F_{\rm tot} - F_{\rm cloud} \\
 & = (F_{\rm cont} + F_{\rm BLR}) - (F_{\rm cont} + f_{\rm cov} \cdot F_{\rm BLR}) ~,
\end{split}
\end{equation}
\noindent
which yields
$$ {\rm LFR} = (1-f_{\rm cov})\cdot \frac{F_{\rm BLR}}{F_{\rm tot}}~.$$

In order to recover $f_{\rm cov}$, we thus need to measure the two quantities $F_{\rm tot}$ and $f_{\rm cont}$ (since $F_{\rm BLR}=F_{\rm tot}-F_{\rm cont}$) in the spectrum at the wavelength of the \ion{C}{i}\,$\lambda1656$ absorption line. The total flux is easy to measure with good precision; however, the continuum flux is more cumbersome as the intrinsic quasar continuum is notoriously difficult to assess. We infer $F_{\rm cont}$ by interpolating a spline locally in the regions around the \ion{C}{iv} emission line (for details see Fig.~\ref{fig:CIV_cont} of the Appendix). Due to the high level of uncertainty in the continuum evaluation, we assign an error of 10 per cent on the continuum flux. This way we obtain $F_{\rm cont}=22.9\pm2.3$ and $F_{\rm tot}=29.3\pm1.5$, both in units of $10^{-16}~{\rm erg}~ {\rm s}^{-1}~{\rm cm}^{-2}~{\rm \AA}^{-1}$.
Given the best-fitting value of the LFR of $16\pm3$~\%, the flux measurements result in a covering fraction of the BLR of $f_{\rm cov} = 27\pm6$~\%.

\section{Dust Extinction}
\label{dust}
In \citet{Krogager2015}, we assigned the reddening to the quasar itself due to the limited data available. However, in this work we have accurate abundance ratios from the DLA and a much extended wavelength coverage from the X-shooter spectrum.
The observed [Fe/Zn] ratio indicates a significant amount of dust in the DLA. First we constrain the dust extinction, $A(V)$, along the line of sight by fitting a quasar template to the observed spectrum. We then compare the best-fitting $A(V)$ to an estimate from the observed depletion ratio of refractory to volatile elements. In the following, we will describe both methods in detail.

\subsection{Template Fitting}
\begin{figure*}
  \includegraphics[width=0.98\textwidth]{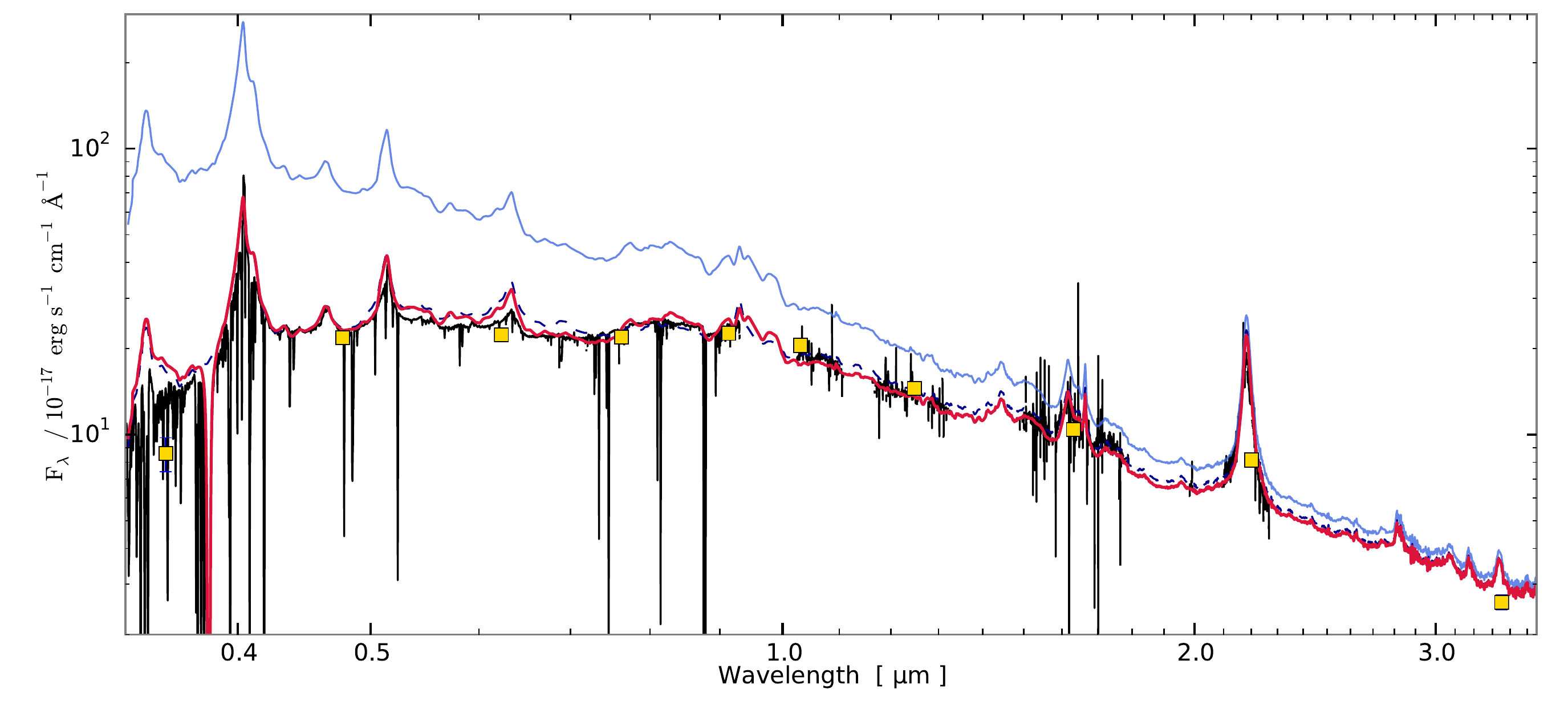}
  \caption{The X-shooter spectrum of quasar J\,2225+0527 shown in black together with the photometry from SDSS, UKIDSS, and WISE band 1 
(shown in yellow squares, the errorbars are too small to be seen). The spectrum in the top panel has been smoothed for presentation purposes.
The telluric absorption regions in the near-IR (at $\sim$1.0, 1.1, 1.5 and 1.9~$\mu$m) have been masked out. 
The blue solid curve shows the unreddened quasar template, and the red solid curve shows the same template assuming the best-fitting dust model with iron emission and Ly$\alpha$ absorption included, see Sect~\ref{dust}. For reference, the blue dashed line shows the best-fitting dust model without including the iron emission template. The largest offset between the red and the dashed line is observed around the \ion{C}{iii}] emission line at $\sim0.6~\mu$m.
\label{fig:spectrum}}
\end{figure*}

By fitting a quasar template to the spectroscopic data, we were able to model the dust extinction along the line-of-sight independently from the depletion analysis. We parametrized our model following the method in \citet{Krogager2015}; However, we expanded the model to contain a variable intrinsic power-law slope, $\beta$, to be able to propagate the correlation between the reddening and the unknown intrinsic slope. In this work, we used the quasar template of \citet[][in preparation]{Selsing2015_prep} and fitted for both the intrinsic quasar dust, ${\rm A(V)_{QSO}}$, and for dust in the DLA, ${\rm A(V)_{DLA}}$. We applied SMC, LMC and LMC2 type extinction curves by \citet{Gordon03} in the fitting routine, and for the quasar dust we furthermore applied a new extinction curve derived explicitly for quasars \citep[see][submitted]{Zafar2015_prep}. We fitted the entire spectrum red-wards of Ly$\alpha$ excluding the regions influenced by the broad emission lines and telluric absorption. Furthermore, we used photometric data from WISE band 1 at $3.5~\mu$m. The WISE bands at longer wavelength were not used since the quasar template does not cover these wavelengths. We did not include optical nor NIR photometric data points since all the available photometry was covered by the spectrum. During the initial fit, we noted an apparent mismatch between the template and the data in the spectral range around the \ion{C}{iii}] emission line of the quasar. We first suspected this to be caused by the 2175~{\AA} bump feature characteristic of the LMC and LMC2 types of extinction curves. However, neither LMC nor LMC2 extinction curves resulted in better fits. Instead we discovered that the mismatch could be explained by variations in the broad emission from \ion{Fe}{ii} and \ion{Fe}{iii}, which form a pseudo-continuum in this wavelength range. We used the template of \citet*{Vestergaard2001} to model the iron emission. The contributions of \ion{Fe}{ii} and \ion{Fe}{iii} were separated into two individual templates, and subsequently smoothed to match the observed width of the broad emission lines of the quasar ($\sim1500$~km~s$^{-1}$). By including the variable iron templates, we obtained a much improved fit, see Fig.~\ref{fig:spectrum}. The best-fitting solution was obtained for the following parameters:
${\rm A(V)_{QSO}}=0.005^{+0.008}_{-0.004}$, ${\rm A(V)_{DLA}}=0.280^{+0.005}_{-0.009}$, and $\beta=0.128\pm0.007$ assuming SMC type dust in the DLA. The results were insensitive to the extinction law in the quasar due to the negligible amount of dust at the redshift of the quasar, also the results were not sensitive to the initial parameter estimate.
The best fit was obtained for an excess of \ion{Fe}{ii} and a deficit of \ion{Fe}{iii} relative to the average quasar template. The errors stated here only represent the statistical errors from the fit. 
We note that a fit to the data without the parameter $A(V)_{\rm DLA}$, i.e., with only dust in the quasar, gives an equally good fit. However, from the high level of depletion ($[{\rm Fe/Zn}]=-1.22$) we have a strong prior indicating dust in the DLA. Moreover, the presence of neutral carbon (due to its ionization potential of 11.26~eV) requires the cloud to be shielded from the background UV radiation. This leads to a lower photo-dissociation rate for dust and molecules. The detection of strong H$_2$ absorption thus provides further evidence for dust in the absorbing medium since H$_2$ forms most effectively on the surface of dust grains (e.g., figure 8 of \citealt{Noterdaeme2008}).

The extinction derived from the fit is strongly correlated with the intrinsic slope of the quasar power-law; this introduces a large systematic uncertainty, since we do not a priori know the intrinsic quasar slope. Previous studies have shown the intrinsic slope of quasars to be normally distributed with a width of $\sim0.2$ \citep[][but see also Krawczyk et al. 2015 and Selsing et al. 2015]{vandenBerk2001}Â \nocite{Krawczyk2015}. By propagating this intrinsic uncertainty on the slope in our fitting, we can estimate the systematic effect on \Avabs. A steeper intrinsic slope (smaller $\beta$) yields a larger extinction. The change in extinction varies approximately as a linear trend with the slope as $\Delta {\rm A(V)_{DLA}} = -0.07\cdot (\Delta\beta/0.02)$, where $\Delta {\rm A(V)_{DLA}}$ and $\Delta\beta$ denote the change in best-fitting extinction and slope, respectively.
We can include this as a systematic uncertainty in our result:
${\rm A(V)_{DLA}} = 0.28\ \pm 0.01|_{\rm stat}\ \pm 0.07|_{\rm sys}$. We note, that in cases of very shallow slopes the extinction becomes very small and these models prefer dust in the quasar. However, such solutions are inconsistent with the broad-band data from WISE at 3.5~$\mu$m and the measure of extinction from the depletion ratio of metals lines in the DLA, see above. Furthermore, the very high luminosity of the quasar (among the top 0.1 per cent brightest quasars at this redshift, $\lambda L_K = 7.9 \times 10^{46}~{\rm erg~s}^{-1}$) renders such models highly unlikely as bright quasars typically have steeper slopes \citep{Davis2007}, which would increase our best-fitting value for \Avabs.

\subsection{Depletion of iron relative to zinc}
Following \citet{Vladilo2006} and \citet{DeCia2013}, we calculated the column density of iron in the dust phase, $\log(\hat N_{\rm Fe})$. This quantity is expected theoretically to correlate with A(V), and we used the empirical relation from \citet{Vladilo2006} to calculate \Avabs\ assuming that the intrinsic abundance ratio of (Fe/Zn) is Solar. Although zinc is usually assumed to be depleted very little onto dust grains, and as such a good estimator of the volatile elements, this is not the case for high metallicity systems. For around Solar metallicity and above, even zinc may show signs of weak depletion \citep{Vladilo2006}. In order to correct for this differential depletion of both iron and zinc, we assumed that the depletion of iron and zinc follow that of the interstellar medium in the Milky Way, see \citet{DeCia2013}. Given the high metallicity ($[{\rm Zn/H}]=-0.09$) inferred from zinc, both the Solar intrinsic abundance ratio and the Galactic depletion pattern are valid assumptions.
We measured a high depletion of $[{\rm Fe/Zn}]=-1.22\pm0.06$, from which we inferred an amount of dust extinction of ${\rm A(V)} =0.39^{+0.13}_{-0.10}$~mag. We note that neglecting differential depletion of zinc relative to iron for such high metallicities would underestimate A(V) by 0.11 mag.
The intrinsic ratio of [Zn/Fe] has been studied extensively \citep[for a recent study, see][]{Berg2015}, and a large spread is generally observed. \citet{Berg2015} show that the intrinsic [Zn/Fe] ratio is most probably {\it lower} than Solar. This would lead to a higher $A(V)$ in this case. Even with a higher than Solar intrinsic ratio of [Zn/Fe] ($+0.2$~dex), the high level of depletion observed in this DLA requires a significant amount of dust in order to reconcile the observed zinc-to-iron ratio with the intrinsic ratio.

\section{The DLA Galaxy Counterpart}
\label{emission}
In order to identify the galaxy associated with the DLA we have searched for stellar continuum emission in broad-band imaging as well as nebular emission lines in the X-shooter spectrum. However, the bright background quasar impedes detection of the faint glow from the DLA galaxy, and it is therefore necessary to subtract as much of the quasar's light as possible. The methods utilized for this subtraction is described below for each dataset.

\subsection{Optical Imaging}
\begin{figure*}
  \includegraphics[width=0.98\textwidth]{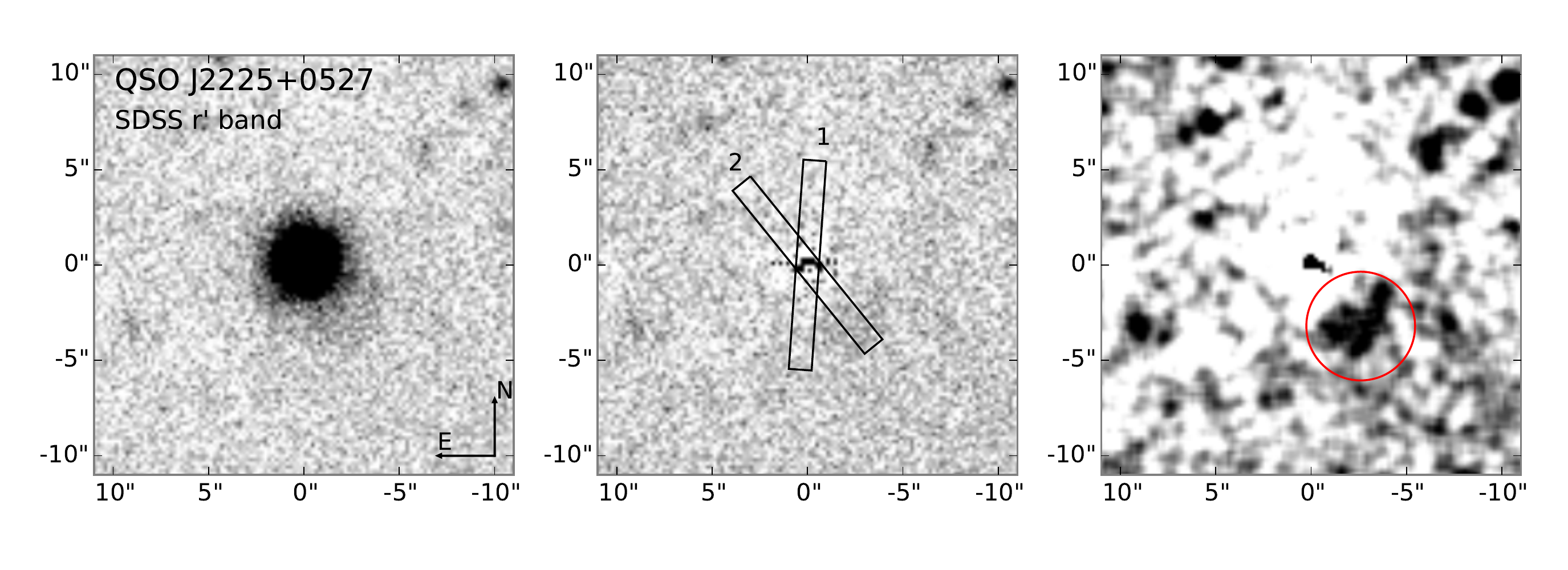}
  \caption{{\it r'}-band image from the Nordic Optical Telescope of the quasar J\,2225+0527. The three panels show (left to right) the combined image, the residuals after subtraction of the point spread function, and the residual image smoothed by a (5$\times$5 pixel$^2$) Gaussian filter with $\sigma_{\rm filter}=2$~pixels. In the middle panel, the two slit positions used for the X-shooter observations are shown as rectangles. In the right panel, an extended excess of emission is visible about 4 arcsec southwest of the quasar (marked by a red circle).   \label{fig:NOT_image}}
\end{figure*}

For the broad-band images, we had to replicate the point spread function (PSF) in order to subtract the point-like quasar emission. In the NOT $r'$-band image, we modelled the PSF by combining bright, though not saturated, stars in the field. The light from the quasar was then subtracted using this empirical PSF. This way we could probe the regions closer to the quasar. However, the central part (within 0\farcs 9) was still not accessible due to strong residuals from the PSF subtraction. The PSF subtracted $r'$-band image is shown in Fig.~\ref{fig:NOT_image}. In the residual image, we observed no emission consistent with that of a DLA galaxy. However, when smoothing the image with a Gaussian filter we detect a fuzzy blob of emission around 4 arcsec South-West of the quasar. This extended emission is consistent with the Ly$\alpha$ nebula observed by \citet{Heckman1991}, and moreover is located at the same spatial position as the extended radio lobes reported by \citet{Barthel1988}. We therefore ascribe this extended emission to the \ion{C}{iii} line of the Ly$\alpha$ nebula, rather than emission from the DLA galaxy.
Hence, the DLA galaxy remains undetected down to a limit of $m_{\rm AB} > 24.79$ mag (3 $\sigma$ limit). Since we were not able to probe the region closest to the quasar we used apertures placed around the quasar at a distance of 1\farcs 9 from the quasar to estimate the flux limit. This corresponds to a physical separation of 16~kpc at the redshift of the DLA. The observed $r'$-band probes flux at $\sim1900$~{\AA} in the rest-frame of the DLA galaxy. This part of the UV continuum is sensitive to recent star formation, and the flux limit thus allows us to constrain the star-formation rate using \citet{Kennicutt1998}. First we applied a correction for Galactic extinction of $A_{r'}=0.347$ from the maps of \citet{Schlafly2011}, and converted the corrected flux to a luminosity given the luminosity distance for $z=2.13$. Finally, we converted the luminosity to star-formation rate (using Chabrier initial mass function) correcting for dust in the DLA assuming an extinction of 1.5~mag at rest-frame 1900~\AA. This way we obtained a 3$\sigma$ limit on star formation rate of ${\rm SFR < 21~M_{\odot}\,yr^{-1}}$, under the assumption that the impact parameter is more than 16~kpc. For smaller impact parameters, the star formation rate may be even higher.

\subsection{Near-infrared Imaging}
\begin{figure}
  \includegraphics[width=0.48\textwidth]{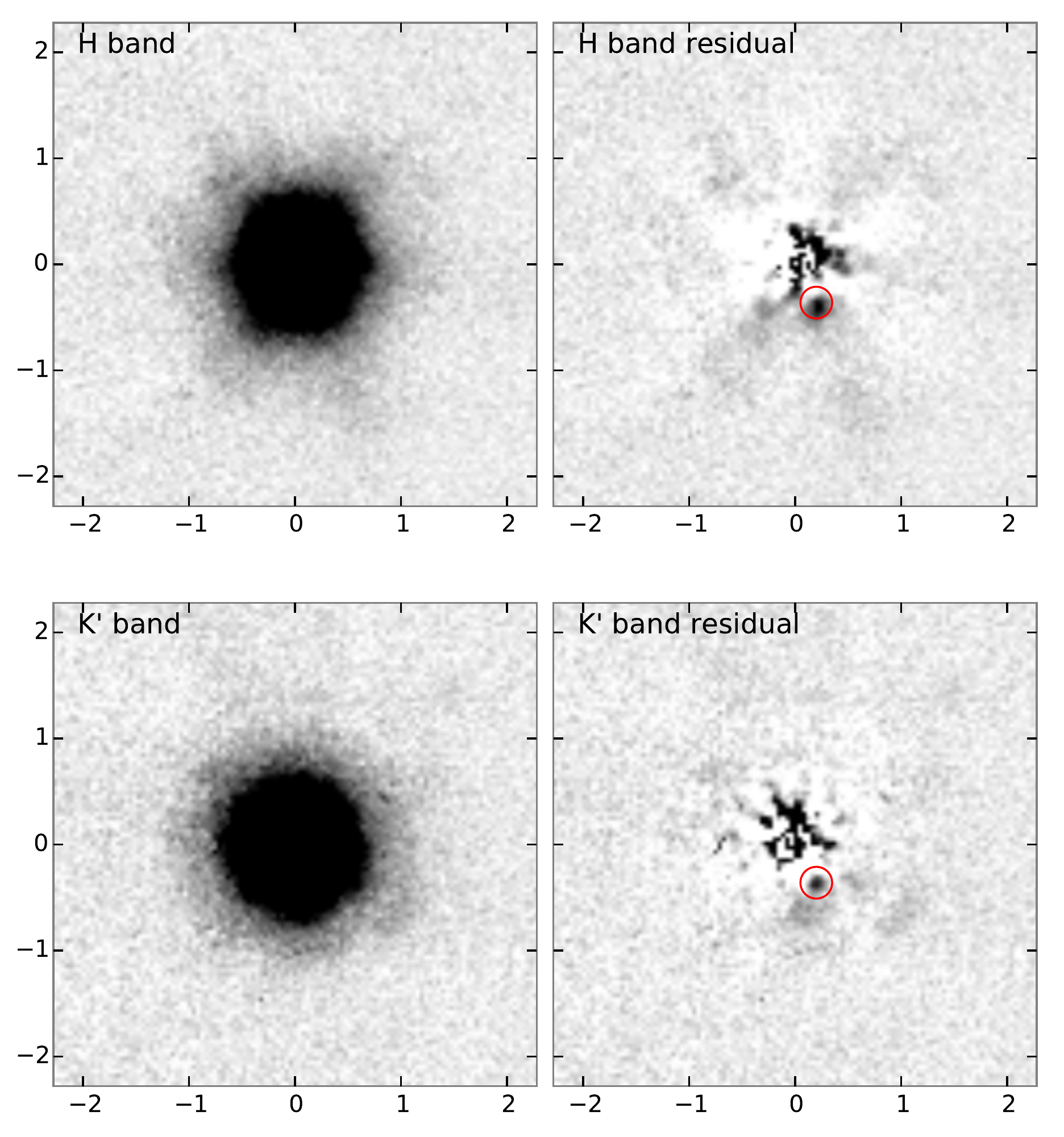}
  \caption{Keck AO images of the quasar J\,2225+0527 in $H$ and $K'$ bands. The left and right panels in each row show, respectively, the image before and after subtraction of a model PSF (see text). Each images are oriented North to the top and East to the left. The red circle in each residual image marks the location of the tentative detection of the emission counterpart of the DLA at redshift $z=2.13$. The radial structure observed in the $H$-band residuals is an artefact of the circular averaging of the PSF. The $J$-band image did not result in a successful PSF subtraction and is therefore not shown. \label{fig:KECK_image}}
\end{figure}

Due to the lack of suitable PSF stars in the small field of view of the Keck AO images we used the quasar itself to make an azimuthally averaged radial PSF profile. We then generated a spherically symmetric PSF from the average radial profile for each image and subtracted the model PSF from the images. In the $H$ and $K'$ images, we observed a small excess flux at roughly 0\farcs5 south of the quasar's centroid, see Fig.~\ref{fig:KECK_image}. The emission was observed at the same location in the two images, which is not expected for a random excess due to a spurious peak in the residuals. We thus regarded this as a tentative detection of the DLA galaxy counterpart. We extracted the flux from the small excess by fitting a 2-dimensional Gaussian model to the excess emission. By using the measured flux of the quasar in UKIDSS photometry as reference, we obtained the following magnitudes for the tentative DLA counterpart: $H=23.6$ and $K'=23.8$ mag, both on the AB system.
The PSF in the $J$ band image was less symmetric than in the other two AO images. This made the PSF modelling more difficult to perform; As a result, we did not obtain a good subtraction of the PSF and no excess emission was observed in the $J$-band at the same position.

\subsection{X-shooter Spectroscopy}
In order to detect weak nebular emission lines from the DLA galaxy in the spectrum, we subtracted the quasar trace from the 2-dimensional spectrum following the method in \citet{Krogager2013}. We did not detect any emission from the nebular lines H$\alpha$, H$\beta$ or [\ion{O}{iii}] at the redshift of the DLA in our X-shooter spectrum. We obtained flux limits for [\ion{O}{iii}] $\lambda 5007$ and H$\alpha$ of $F_{\rm O\,III} < 5.3 \times 10^{-17}\,{\rm erg\, s^{-1}\, cm^{-2}}$ (3$\sigma$) and $F_{\rm H\alpha} < 16.5 \times 10^{-17}\,{\rm erg\, s^{-1}\, cm^{-2}}$, respectively. The spatial extent of the aperture used for the flux limit estimation was determined from the width of the spectral trace (FWHM = 5 pixels at the location of H$\alpha$). We used an aperture size of twice the FWHM (1.5 arcsec) centred on the position of the quasar. In the spectral direction, we used the velocity width of the absorption lines (330~km~s$^{-1}$) centred around the absorption systemic redshift $z_{\rm sys}=2.132489$. This provides a conservative upper limit which takes into account the fact that we do not know the exact emission redshift, however, the emission redshift is probably within the velocity spread observed in the different absorption components. Also, we assumed that the emission is at very small impact parameter, which seems plausible given the hints from the Keck imaging described above.
Unfortunately the H$\alpha$ line is located in a region of Telluric absorption lines. The flux limit is therefore not very constraining; However, it still allows us to put a limit on the star formation rate by using the relation from \citet{Kennicutt1998}. This yields an upper limit on the star formation rate of ${\rm SFR_{H \alpha}} < 29~{\rm M}_{\odot}~{\rm yr}^{-1}$, assuming Chabrier IMF and correcting for dust.

\section{Discussion}
\label{discussion}

The quasar observed in this work, although identified in previous radio catalogs, was missing in the SDSS quasar database due to significant amounts of dust causing the optical colours of the quasar to appear stellar-like in the SDSS colour spaces. However, since the quasar is a radio-loud quasar it would be expected to appear in the SDSS database. The reason why this quasar does not appear in SDSS is that the cross matching of SDSS and the FIRST radio catalog was performed in 2008, whereas Q2225+0527 was not observed until 2009. It is thus an unfortunate coincidence of various factors that resulted in the misclassification of Q2225+0527 in the SDSS database.

The foreground DLA at $z=2.13$ was first reported by \cite{Barthel1990} and later by \cite*{Junkkarinen1991} and \cite*{Ryabinkov2003}; However, the dust reddening towards the quasar has so far not been studied in detail. Based on the detection of \ion{C}{i} and H$_2$ together with high depletion from [Zn/Fe], we had strong evidence for dust in the DLA. This was further bolstered by our spectral fitting. The best-fitting model had all the dust in the DLA. We found the DLA to be of Solar metallicity with $[{\rm Zn/H}]=-0.09\pm0.05$ and we inferred a dust-to-gas ratio of ${\rm A(V)}/N({\rm H\, {\textsc i}}) = 5.5\times 10^{-22}~{\rm mag~cm^2}$ consistent with that of the local ISM \citep[$4-6\times 10^{-22}~{\rm mag~cm^2}$;][]{Liszt2014}. This value of dust-to-gas ratio is much higher than the average DLA (${\rm A(V)}/N({\rm H {\textsc i}}) = 2-4\times 10^{-23}~{\rm mag~cm^2}$; \citealt{Vladilo2008}).

\begin{figure}
  \includegraphics[width=0.48\textwidth]{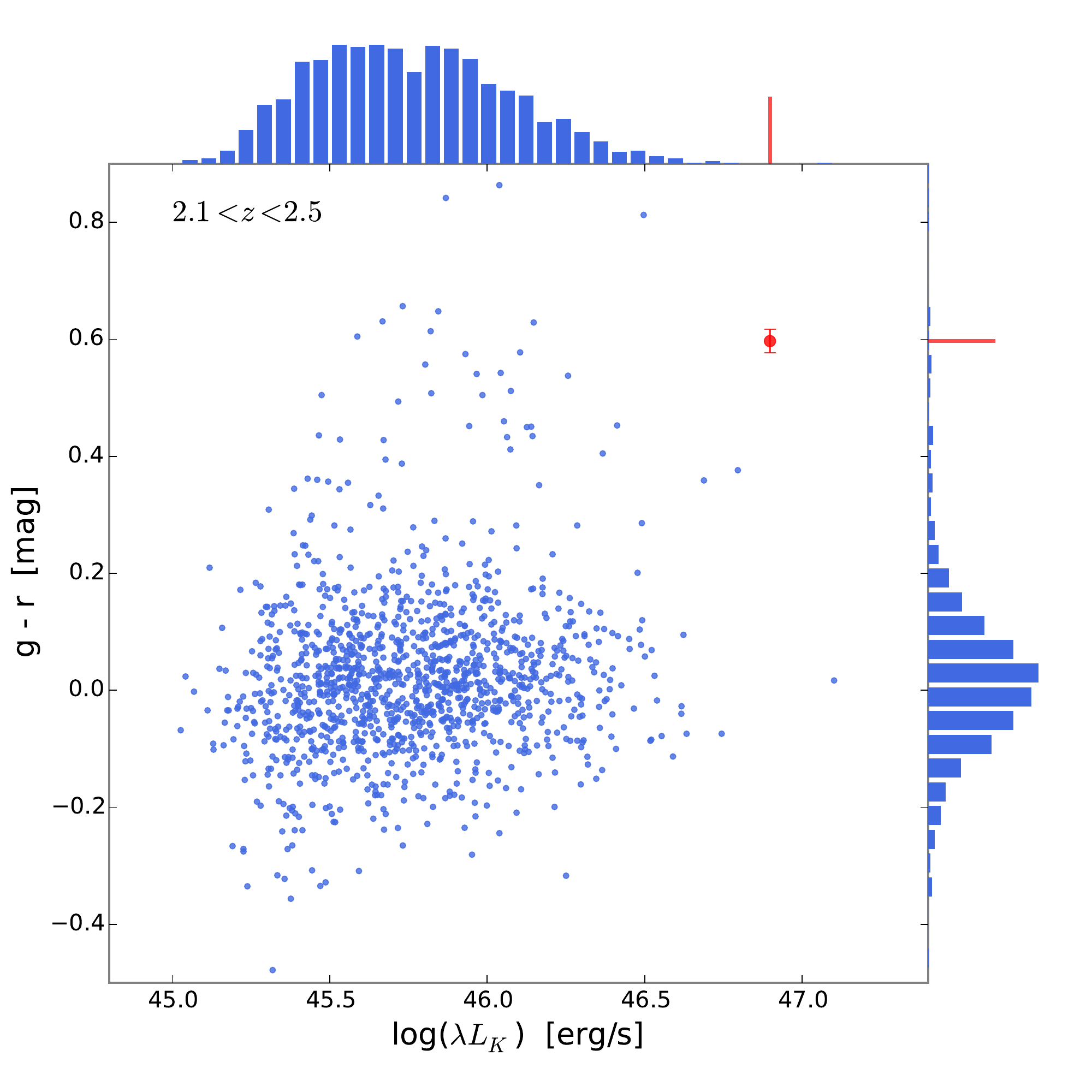}
  \caption{Distribution of SDSS $g-r$ color and $K$-band luminosity for quasars with redshifts from $2.1<z<2.5$ (blue points and histograms) selected from the sample of \citet{Peth2011}. The quasar presented in this work is shown in red. All magnitudes have been corrected for Galactic extinction of $E(B-V)=0.132$.
  \label{fig:grK}}
\end{figure}

\begin{figure}
  \includegraphics[width=0.48\textwidth]{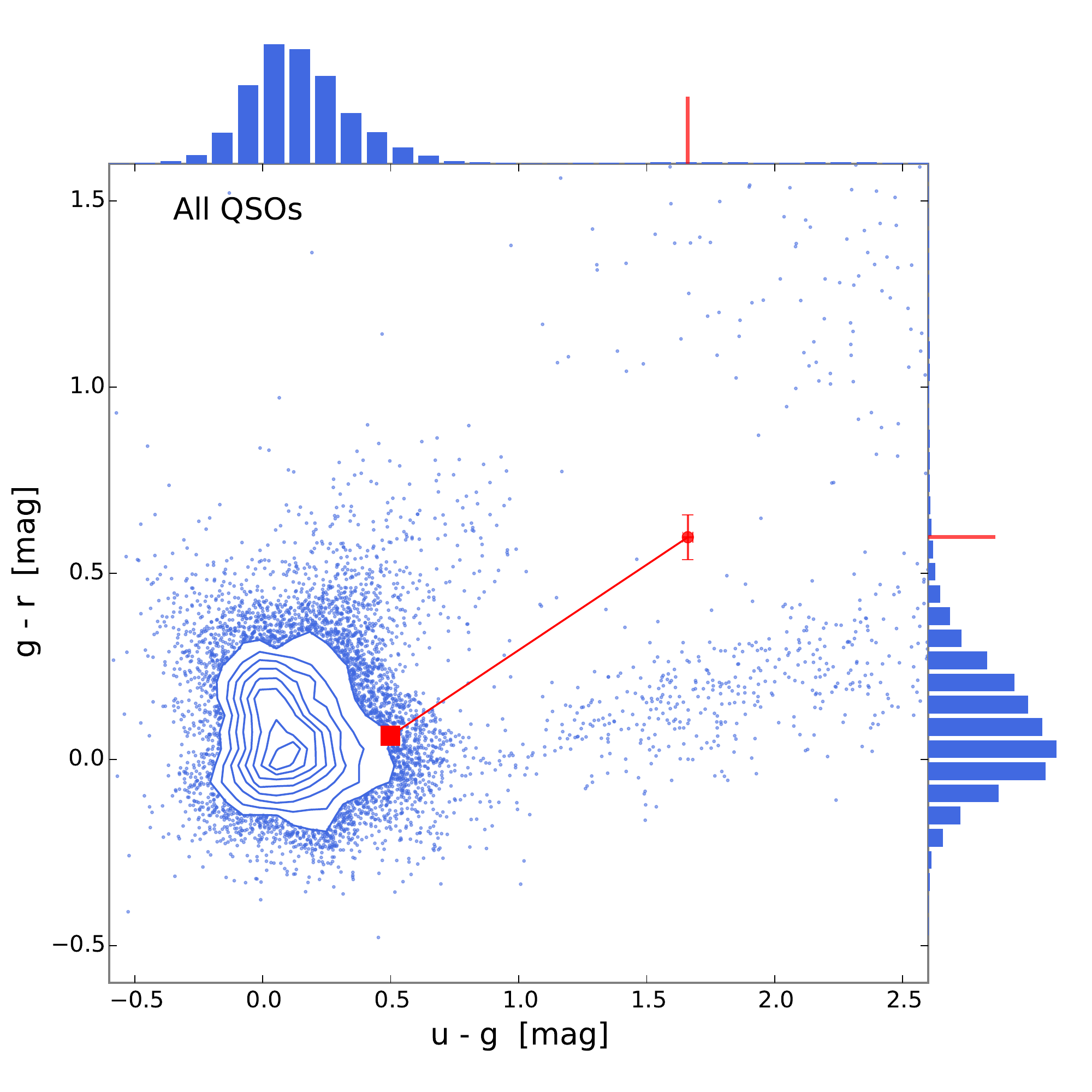}
  \caption{Distribution of SDSS $u-g$ and $g-r$ colors for quasars
  (blue contours and points) selected from the sample of \citet{Peth2011}.
  The red dot marks the observed colours of the quasar, Q2225, presented
  in this work. The red square connected to the red dot by the line marks
  the colours of Q2225 after correcting for the foreground reddening from
  the DLA. The intrinsic colours of the quasar are thus consistent with
  the bulk of quasars identified in SDSS, though slightly redder than the
  average. This is consistent with the redder than average intrinsic slope
  inferred from the spectral fitting.
  All magnitudes have been corrected for Galactic extinction of $E(B-V)=0.132$.
  \label{fig:ugr}}
\end{figure}

This shows that dusty DLAs indeed do exist and that they are, to some degree, missed in optically selected quasar samples. The quasar studied here is intrinsically very bright and red (among the top 0.1 per cent most luminous and top 1 per cent most red quasars at these redshifts, see Figures \ref{fig:grK} and \ref{fig:ugr}). A bias against dusty DLAs, and absorption systems in general, is therefore not only limited to the faint end of the quasar luminosity function. The bright quasars are also affected due to the reddening caused by dust allowing the quasars to escape detection by optical colour criteria. The actual colour criteria used by SDSS are hard to visualize, since the algorithm works in a multi-dimensional colour-space \citep{Richards2002}. However, in order to visualize the effect of dust reddening we show the $u-g$ and $g-r$ colour-colour diagram in Figure~\ref{fig:ugr}. The contours and small blue points indicate SDSS quasars, showing that they mostly cluster in the region around $(0,0)$. The individual points at redder colours are added to the SDSS sample by various other criteria such as radio selection \citep[for comparison, see][]{Richards2001, Richards2002}. The red dot marks the observed colours of the quasar in this work and its intrinsic colours after correcting for the DLA dust (the red square connected to the red dot by a line). The quasar thus falls in the region of optically selected quasars if there had been no dust in the DLA. Owing to its very bright radio emission, the quasar presented here was identified in radio surveys. However, only about 10 per cent of quasars are radio loud \citep{Balokovic2012}. Hence, more dusty absorbers might be missing toward red, radio quiet quasars.
The DLA presented here is similar to the recent sample of neutral carbon absorbers by \citet{Ledoux2015} both in terms of \ion{C}{i} abundance and dust extinction. These authors also argue that the most dusty absorption systems are missing in current samples selected using optical criteria.

The impact of this bias, i.e., the frequency of dusty DLAs, is not a large effect when seen in terms of the overall number of DLAs. The criteria used in the HAQ survey \citep{Krogager2015} select around 2 per cent of the general quasar population (known in SDSS data release 10). Of the quasars identified in the HAQ survey, about 2 per cent have dusty foreground absorbers. Hence, only a small number of absorbers are missed. However, these absorbers are primarily metal-rich and will contribute to a significantly larger fraction of the metals, since the bulk of DLAs have very low metallicities. This is consistent with what has been found in radio surveys of DLAs \citep{Ellison2001a, Jorgenson2006}, i.e., dust biasing does not have a strong effect on the bulk of the DLA population.
We furthermore caution the reader that the numbers presented here are only rough approximations and a more detailed investigation of the selection effects is still needed.

%
%
% DISCUSSION OF PARTIAL COVERAGE AND DUST.
% DETERMINE BLR SIZE FROM CLOUD SIZE!
\subsection{Partial Coverage and Size of the BLR}
In Sect.~\ref{partial_coverage}, we derived the line flux residual from the \ion{C}{i} absorption lines. If we interpret the LFR as a geometric effect caused by partial coverage of the background emitting source (in this case the broad line region of \ion{C}{iv}) then the LFR can be converted to a covering fraction. Other explanations for the LFR would be gravitational lensing of the quasar or a double quasar, see the discussion in Balashev et al. (2011). As argued in Sect.~\ref{partial_coverage} we favour the geometric effect since the other scenarios require a somewhat unlikely conspiracy of alignments. However, with the current data, we cannot exclude the other scenarios and we caution the reader that the following exercise is merely suggestive.

Given the covering fraction of 27$\pm$6 per cent of the BLR, inferred in Sect.~\ref{partial_coverage}, we can calculate the size of the BLR assuming a simple spherical geometry of both the absorbing cloud and the BLR, as depicted in the schematic view of Fig.~\ref{fig:CI_overview}. A prerequisite for this calculation is the physical size of the absorbing cloud, which we can estimate from the fine structure levels of C\,{\sc i}. These fine structure levels are excited by the background radiation field and by collisions with other species predominantly hydrogen (both atomic and molecular). The population of the various levels will therefore depend on the density and temperature of the gas. In the following, we assume that neutral carbon arises from the cold self-shielding cloud where molecular hydrogen is the primary collisional partner. We then use the numerical code {\sc popratio} (Silva \& Viegas 2001) to calculate the population of various fine-structure levels for a grid of temperatures (50$-$500 K) and densities (1$-$1000~cm$^{-3}$). Assuming that the column densities derived in Sect.~\ref{partial_coverage} trace the physical densities, i.e., $N({\rm C}\, \textsc{i})/N({\rm C}\, \textsc{i}^{*})=n({\rm C}\, \textsc{i})/n({\rm C}\, \textsc{i}^{*})$, we can then calculate the allowed range in temperature and density ($n_{\rm H_2}$). The temperature and density are highly degenerate and the temperature remains unconstrained, though typical values for the cold, neutral phase lie in the range of $50-200$~K. However, we obtain a range of allowed densities from 90 to 400 cm$^{-3}$ with a most probable density around 100~cm$^{-3}$ for $T=100$~K.

With the physical density of H$_2$ at hand, we can infer a physical length scale of the small absorbing cloud from the ratio of column density to volumetric density: $l_{\rm cloud} = N/n \approx 0.1$~pc. We assume that the cloud is spherically symmetric with radius $r_{\rm cloud}=l_{\rm cloud}/2$. In projection as shown in the top right panel of Fig.~\ref{fig:CI_overview}, the system will therefore appear like two overlapping disks with an offset between the two centres, $\Delta l$. Since we do not know this projected separation we calculate a grid of various separations and radii of the BLR. For each grid point, we can then evaluate the covering fraction of the BLR by the small cloud of radius $r_{\rm cloud} = 0.05$~pc. Since the quasar and absorbing cloud are independent the projected separation between the centres of the two disks will be randomly distributed. For the spherically symmetric geometry that we assume, the probability of a given $\Delta l$ will depend linearly on the separation $p(\Delta l) \propto \Delta l$, i.e., there is zero probability of a perfect alignment. We furthermore require $\Delta l \le r_{\rm cloud}$ since the continuum source (the accretion disk) in the centre of the BLR must be covered by the absorbing cloud. These two requirements serve as our prior on the separation: $p(\Delta l) = 2\Delta l/r_{\rm cloud}^2$ for $\Delta l\le r_{\rm cloud}$ and $p(\Delta l)=0$ for $\Delta l>r_{\rm cloud}$.
Given the measurement of the covering fraction we can convert the 2-dimensional grid of covering fraction as a function of $R_{\rm BLR}$ and $\Delta l$ to a likelihood function. By multiplying the likelihood function with the prior on $\Delta l$ and marginalizing over all separations, we obtain the most probable radius of the \ion{C}{iv} broad line region, $R_{\rm BLR}=0.1$~pc. However, given the large uncertainties on the density, temperature, and size of the cloud, the size of the BLR is not constrained to better than a factor of $\sim$2, i.e., $R_{\rm BLR} \approx 0.05 - 0.2$~pc.

From the observed correlation between the physical size of the broad line region and luminosity, we can independently estimate the radius of the BLR. We use the relation for \ion{C}{iv} from \citet{Kaspi2007}, who quantify the relation in terms of the luminosity at 1350~\AA. In the X-shooter spectrum, we measure $\lambda L_{\lambda}(1350)=3.48\times 10^{46}$~erg~s$^{-1}$, which corresponds to $R_{\rm BLR}=0.17$~pc with an uncertainty of 54\,\% dominated by the scatter in their calibration. This is well within the uncertainties on both the radius--luminosity relation and our estimate. The good agreement between the two size estimates provides evidence for the partial coverage due to the small projected size of the cloud compared to the BLR, despite the simple geometry we assumed.

%
%   SECTION FROM PALLE
%
%
\subsection{Stellar Mass of the DLA Galaxy}
The velocity structure of the absorbing gas contains 10 individual
components spread over a total range of 397~{\kms}. Such a wide
span of velocities is in fact quite common for absorbers with a similar
high metallicity \citep{Ledoux2006}, and it is believed to be related to
the deep potential well of the massive dark matter haloes that host
high metallicity galaxies \citep{Moller2013, Neeleman2013, Arabsalmani2015}.
In Sect.~\ref{absorption}, we quantified the velocity width of the system
using $\DeltaÂ V_{90}$ for which we obtain a value of 331~km~s$^{-1}$.
The resolution of X-shooter at the observed wavelength of the line is
27.2~km~s$^{-1}$, and we use the method described by Arabsalmani et al. (2015)
to correct the $\DeltaÂ V_{90}$ measured and obtain the intrinsic value
$\DeltaÂ V_{90} =  329~{\rm km~s}^{-1}$.

\begin{figure}
  \includegraphics[width=0.48\textwidth]{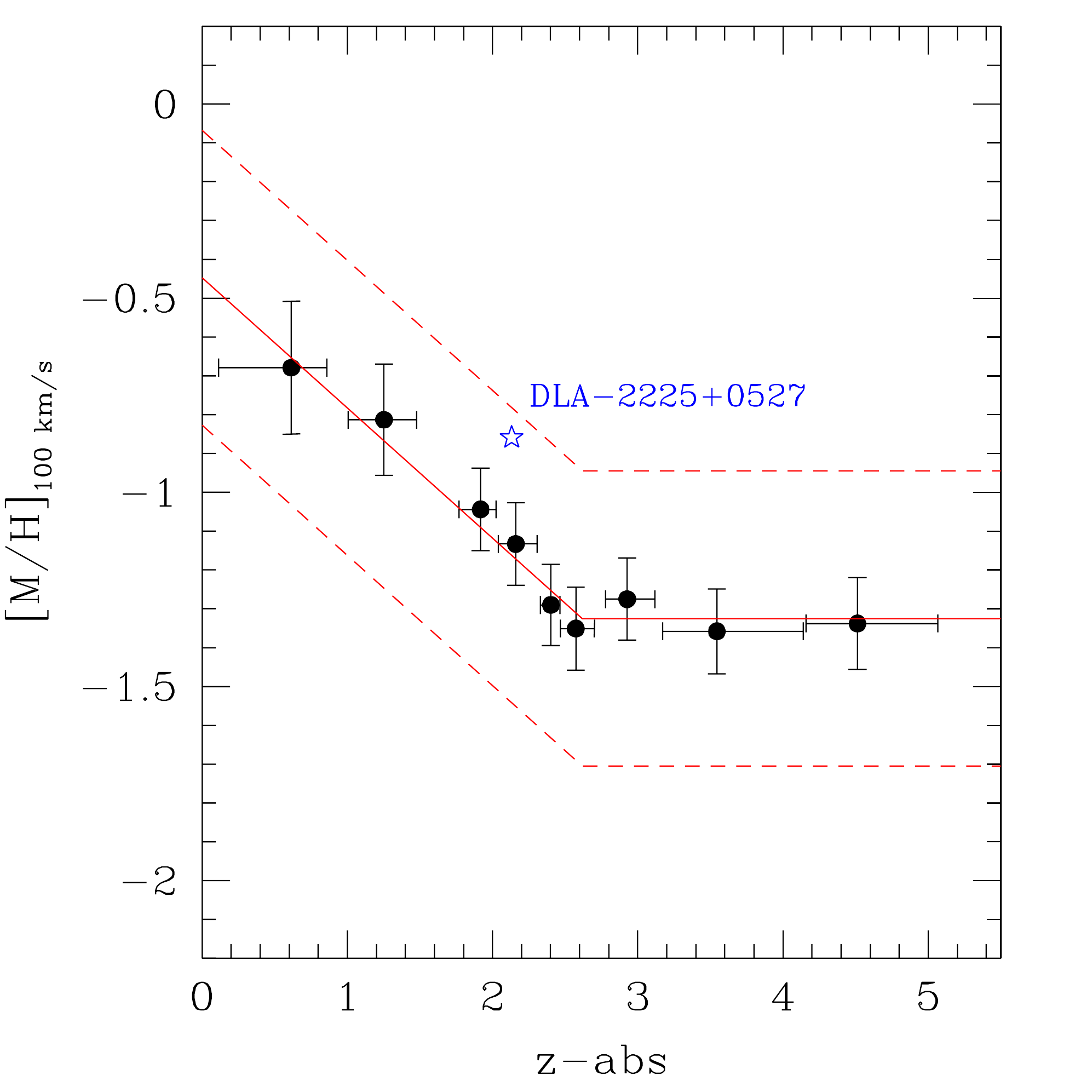}
  \caption{Comparison to the redshift evolution
of the global metallicity-$\Delta V_{90}$ relation of DLAs.
The quantity [M/H]$_{\rm 100~km/s}$ denotes the metallicity
normalization of the relation at 100~${\rm km~s^{-1}}$.
Black dots and full red line marks the binned data
and simple linear fit, red dashed lines mark the
1 $\sigma$ internal scatter of the relation (from
\citet{Moller2013}). DLA-2225+0527 is seen to lie well
within the general distribution of intervening DLAs.
  \label{fig:metal_redshift}}
\end{figure}

The $\DeltaÂ V_{90}$ has been shown to correlate with metallicity, a
correlation which depends on the redshift \citep{Ledoux2006}. The
evolution of the zero-point of the relation is shown in Fig.~\ref{fig:metal_redshift}
(black points and full red line from \citet{Moller2013}). The $1\sigma$
of the intrinsic scatter of the relation is marked by the dashed red
lines, and the DLA (blue star) is seen to lie well within this scatter.
\citet{Christensen2014} confirmed the expectation that the relation in
reality is a mass-metallictiy relation, and using the prescription from
\citet{Christensen2014}, and [M/H] = $-$0.1, we predict a stellar mass of
$\log({\rm M}_{\star}/{\rm M}_{\odot}) = 10.78\pm0.55$ which is among the few most
massive DLA galaxies known at any redshift.
We note that because we do not know the impact parameter for this
galaxy we use the average value 0.44 for $C_{\rm [M/H]}$. In case the
true impact parameter is very small, $C_{\rm [M/H]}$ would be somewhat
smaller and likewise then the stellar mass (as discussed below).

Comparing to the three known DLA galaxies at similar redshifts \citep[][Tables 2 and 3]{Christensen2014}, those have masses computed from the same
formula in the range $\log({\rm M}_{\star}/{\rm M}_{\odot}) = 9.9-11.0$,
and rest-frame optical AB magnitudes of 23.6, 24.4, and 25.2.
In case we assume that the DLA
galaxy in our case is perfectly aligned with the quasar (i.e., impact
parameter of zero) then we can use equation (3) from Christensen et al.
and obtain the lowest possible stellar mass of $\log({\rm M}_{\star}/{\rm M}_{\odot}) = 10.00$.
This is still a very high stellar mass for a DLA galaxy, but the
galaxy would remain hidden under the glare of the quasar.

We searched for the emission counterpart of the DLA both in the X-shooter spectrum and in imaging. From the spectrum we only obtained upper limits on the [\ion{O}{iii}] and H$\alpha$ line fluxes, and the $r'$-band image only provided a flux limit for impact parameters larger than 16~kpc. Interestingly, we observed tentative signs of the emission counterpart in the near-infrared AO images in the $H$ and $K'$ bands. Although this is highly uncertain given the relatively simple PSF subtraction, it is however reassuring that the excess emission is observed at the same spatial location in the two images. We can thus speculate that the DLA is associated with the excess emission observed in the near-infrared images at an impact parameter of roughly 4~kpc. The inferred $H$-band magnitude of the emission counterpart would then be 23.6 mag which is in good agreement with the observed fluxes in the sample of Christensen et al. (2014).
In previous studies of emission counterparts of DLAs, we inferred star formation rates of 13 and 27~M$_{\odot}~{\rm yr}^{-1}$ \citep{Krogager2013, Fynbo2013b} for DLAs with slightly lower metallicities of $[{\rm Zn/H}]\approx -0.5$. The upper limit on the star formation rate of ${\rm SFR}<29~{\rm M_{\odot}~yr}^{-1}$ derived in Sect.~\ref{emission} is therefore consistent with previous detections.

\section{Summary}
In this work, we presented imaging and spectroscopic observations of the quasar J\,2225+0527 and found that the DLA at $z=2.13$ has high metallicity (consistent with Solar), high level of depletion (inferred from [Fe/Zn]), and strong molecular hydrogen absorption. Moreover, we detected neutral carbon absorption lines showing signs of partial coverage due to the projected size of the absorbing cloud being small compared to the background source. By deriving the covering fraction and a rough size estimate of the absorbing cloud, we were able to infer the size of the broad line region of \ion{C}{iv} in the quasar to be $0.05-0.2$~pc. This is consistent with the estimate from the radius-luminosity relation from \citet{Kaspi2007} to within the $1\,\sigma$ scatter of 54 per cent of their relation.

From the depletion ratio, we estimated the amount of dust in the DLA to be $\rm A(V)=0.39^{+0.13}_{-0.10}$~mag. Spectral modelling of the quasar resulted in a best-fitting A(V) of ${\rm A(V)_{DLA}}=0.28$~mag with no dust in the quasar itself. We therefore concluded that the reddening of the quasar is caused by dust located in the DLA, and that this reddening caused the quasar to be misclassified as a star in SDSS.

We observed tentative evidence for an emission counterpart to the DLA in the near-infrared $H$ and $K'$ band images at an impact parameter of roughly 0\farcs5 or 4~kpc. The flux of the tentative DLA galaxy in the $H$-band was estimated to be roughly 23.6~AB~mag. This corresponds to a stellar mass of roughly $10^{9}-10^{10}~{\rm M_{\odot}}$ consistent with the estimate from the mass-metallicity relation of \citet{Christensen2014}.

The quasar studied in this work, although identified previously in radio surveys, was misidentified as a star in SDSS due several factors. Had this been a more typical quasar in terms radio-loudness, the quasar would have been missed by the optical criteria utilized by SDSS and furthermore missed by radio surveys.
Therefore, enriched and dusty galaxies detected in absorption towards quasars are underrepresented in the current optically selected samples. This has important implications for the mapping of chemical enrichment, as it introduces a bias against the most enriched systems at high redshift.
Although a small effect in terms of the number of missing absorbers, the bias introduced in the metallicity distribution as well as the cold gas fraction of DLAs still needs to be determined quantitatively.

\section*{Acknowledgments}
We thank Marianne Vestergaard for insightful discussion and
for sharing her iron emission template with us.
We thank the ESO Director General
for allocating our observing program under the director's discretionary time.
The Dark Cosmology Centre is funded by the DNRF. The research leading to these
results has received funding from the European Research Council under the
European Union's Seventh Framework Program (FP7/2007-2013)/ERC Grant agreement
no. EGGS-278202.
This research has made use of the Keck Observatory Archive (KOA),
which is operated by the W. M. Keck Observatory and the NASA
Exoplanet Science Institute (NExScI), under contract with
the National Aeronautics and Space Administration.

\appendix

\section{Absorption Lines}

The full list of identified absorption lines belonging to the DLA is given in Table~\ref{full_linelist}. We further quote the equivalent widths for each line.

\begin{table}
\caption{List of absorption lines from the $z=2.13249$ DLA. \label{full_linelist}}
\begin{center}
\begin{tabular}{l r}
\hline
\hline
Transition & EW$^{(1)}$ \\
\hline

\ion{Fe}{ii},$\lambda1144$	&			0.623$\pm$0.009 \\
\ion{S}{ii},$\lambda1250^{(2)}$	&			$<$0.392$\pm$0.007 \\
\ion{S}{ii},$\lambda1253$	&			0.339$\pm$0.005 \\
\ion{S}{ii},$\lambda1259$ + \ion{Si}{ii} + \ion{Fe}{ii},$\lambda1260$	&	2.294$\pm$0.006 \\
\ion{O}{i},$\lambda1302$ +  \ion{Si}{ii},$\lambda1304$		&	3.157$\pm$0.004 \\
\ion{C}{ii},$\lambda1334$ + \ion{C}{ii}$^*,\lambda1335$		&	2.243$\pm$0.006 \\
\ion{Ni}{ii},$\lambda1370$	&        	0.105$\pm$0.005 \\
\ion{Ni}{ii},$\lambda1454$	&        	0.045$\pm$0.004 \\
\ion{Al}{ii},$\lambda1670$	&            1.967$\pm$0.004 \\
\ion{Ni}{ii},$\lambda1709$	&        	0.074$\pm$0.005 \\
\ion{Ni}{ii},$\lambda1741$	&        	0.119$\pm$0.004 \\
\ion{Si}{ii},$\lambda1526$	&            1.662$\pm$0.006 \\
\ion{Fe}{ii},$\lambda1608$	&            0.754$\pm$0.004 \\
\ion{Si}{ii},$\lambda1808$	&            0.184$\pm$0.008 \\
\ion{Co}{ii},$\lambda1941$	&			$<$0.018	 ($3\,\sigma$)\\
\ion{Co}{ii},$\lambda2012^{(2)}$	&            $<$0.04$\pm$0.02 \\
\ion{Mg}{i} + \ion{Zn}{ii},$\lambda2026$	&	0.276$\pm$0.006 \\
\ion{Cr}{ii},$\lambda2056$	&            0.055$\pm$0.007 \\
\ion{Cr}{ii} + \ion{Zn}{ii},$\lambda2062$	&	0.137$\pm$0.006 \\
\ion{Fe}{ii},$\lambda2249$	&            0.042$\pm$0.005 \\
\ion{Fe}{ii},$\lambda2260$	&            0.086$\pm$0.005 \\
\ion{Fe}{ii},$\lambda2344$	&         	1.899$\pm$0.005 \\
\ion{Fe}{ii},$\lambda2374$	&            0.946$\pm$0.004 \\
\ion{Fe}{ii},$\lambda2382$	&         	2.682$\pm$0.003 \\
\ion{Mn}{ii},$\lambda2576$	&            0.160$\pm$0.004 \\
\ion{Fe}{ii},$\lambda2586$	&         	1.741$\pm$0.004 \\
\ion{Mn}{ii},$\lambda2594$	&            0.144$\pm$0.004 \\
\ion{Fe}{ii},$\lambda2600^{(2)}$	&         	2.973$\pm$0.003 \\
\ion{Mn}{ii},$\lambda2606^{(2)}$	&         	0.278$\pm$0.003 \\
\ion{Mg}{ii},$\lambda2796$	&			4.759$\pm$0.005 \\
\ion{Mg}{ii},$\lambda2803$	&			4.350$\pm$0.005 \\
                            &                           \\
\ion{Mg}{i},$\lambda2852$	&      		0.909$\pm$0.005 \\
\ion{Cl}{i},$\lambda1347$	&        	0.060$\pm$0.005 \\
\ion{C}{i} + \ion{C}{i}$^*$ + \ion{C}{i},$^{**}\lambda1328$	&		0.231$\pm$0.004 \\
\ion{C}{i} + \ion{C}{i}$^*$ + \ion{C}{i},$^{**}\lambda1560$	&		0.216$\pm$0.004 \\
\ion{C}{i} + \ion{C}{i}$^*$ + \ion{C}{i},$^{**}\lambda1656$	&		0.390$\pm$0.003 \\
\ion{N}{i},$\lambda1199$ + \ion{N}{i},$\lambda1200$	&	2.228$\pm$0.010 \\
\ion{N}{i},$\lambda1134.4$	&      		1.451$\pm$0.009 \\
                            &                           \\
\ion{C}{iv},$\lambda1550,1548$	&		2.924$\pm$0.007 \\
\ion{Al}{iii},$\lambda1854$		&		0.559$\pm$0.009 \\
\ion{Al}{iii},$\lambda1862$		&		0.392$\pm$0.009 \\
\ion{Si}{iv},$\lambda1393$	    &  		1.041$\pm$0.006 \\
\ion{Si}{iv},$\lambda1402$	    &  		0.645$\pm$0.006 \\

\hline
\end{tabular}
\end{center}

$^{(1)}$ Rest-frame equivalent width in units of $\mathrm{\AA}$.\\
$^{(2)}$ Contaminated line.

\end{table}

\newpage

\section{Continuum Determination}
\label{continuum}
\begin{figure*}
  \includegraphics[width=0.98\textwidth]{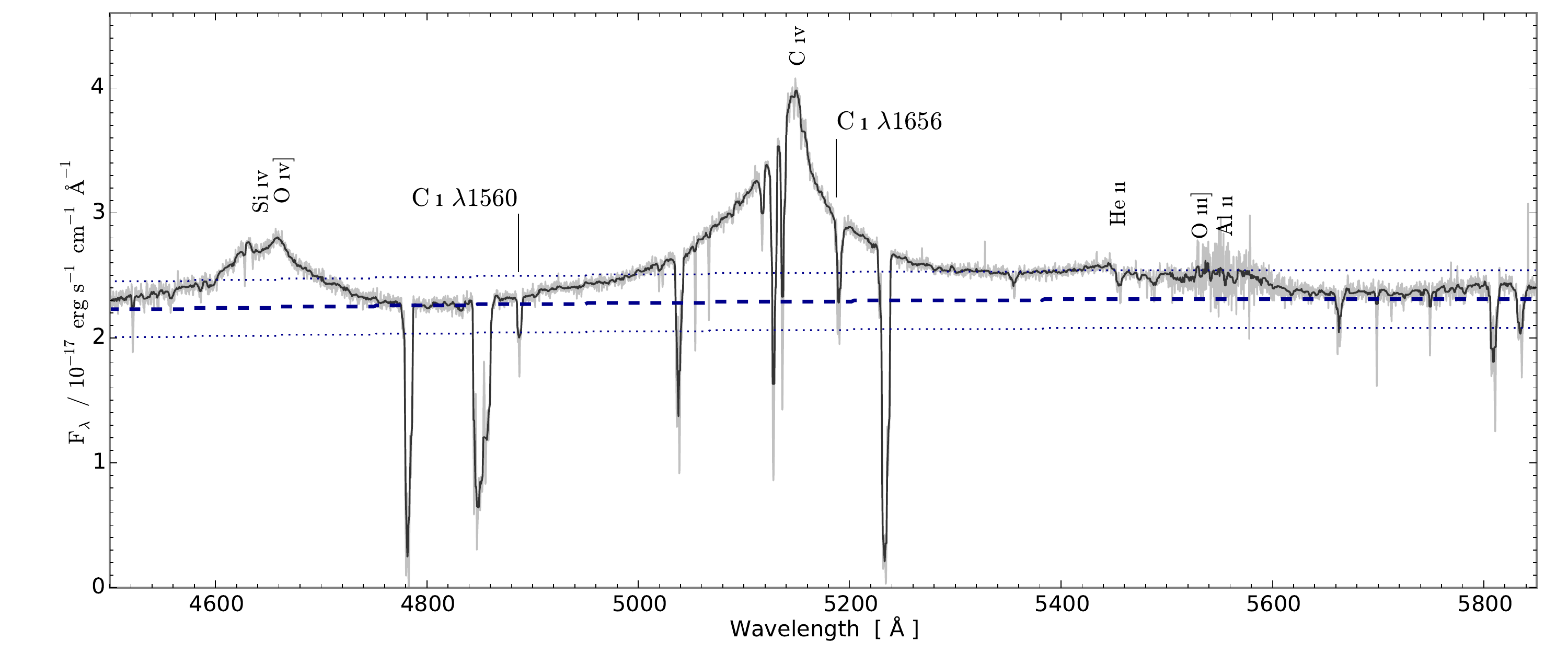}
  \caption{Spectrum of QSO J\,2225+0527 around the \ion{C}{iv} emission line. The spectrum has been smoothed by a 21 pixel median filter for clarity (the dark grey line). The lighter grey underlying line indicates the spectrum before smoothing. The two absorption line complexes from \ion{C}{i} are marked. The estimated quasar continuum is shown by the dashed line. The conservative uncertainty of 10 per cent is shown by the dotted lines. The vertical labels indicate the strongest emission lines from the quasar.
  \label{fig:CIV_cont}}
\end{figure*}

The continuum flux of the quasar around the \ion{C}{iv} line is obtained by interpolation of relatively emission-line free regions. This is however very uncertain, as many broad and weak lines can blend together, which may create a so-called {\it pseudo continuum}. For this reason we assume a conservative uncertainty of 10 per cent for the placement of the continuum. The estimated continuum is shown in Fig.~\ref{fig:CIV_cont}.

\def\aj{AJ}
\def\araa{ARA\&A}
\def\apj{ApJ}
\def\apjl{ApJ}
\def\apjs{ApJS}
\def\apss{Ap\&SS}
\def\aap{A\&A}
\def\aapr{A\&A~Rev.}
\def\aaps{A\&AS}
\def\mnras{MNRAS}
\def\memras{MmRAS}
\def\nat{Nature}
\def\pasp{PASP}
\def\aplett{Astrophys.~Lett.}

\bibliographystyle{mn}

\label{lastpage}

\end{document}

%% file: metal_table.tex
\begin{table*}
\caption{Column densities for individual components of the Voigt profile fit to eight elements.
		  All entries below are given as $\log\left(N/{\rm cm}^{-2}\right)$. \label{tab:fit_details}}
\begin{center}
\begin{tabular}{ccccccccc}
\hline
\hline
$v_{\rm rel}\,^{(a)}$ &  $b$ &   \ion{Si}{ii}   &   \ion{S}{ii}    &   \ion{Cr}{ii}   &   \ion{Mn}{ii}   &   \ion{Fe}{ii}   &   \ion{Ni}{ii}   &   \ion{Zn}{ii}  \\
${\rm km~s^{-1}}$ & ${\rm km~s^{-1}}$ &  &  &  &  &  &  & \\
\hline
    $-183$    & $32$ &      \ldots      &      \ldots      &      \ldots      & $11.49 \pm 0.26$ & $13.50 \pm 0.03$ &      \ldots      &      \ldots     \\
    $-146$    & $11$ & $14.30 \pm 0.17$ & $14.45 \pm 0.07$ &      \ldots      & $11.38 \pm 0.25$ & $13.69 \pm 0.02$ & $12.93 \pm 0.15$ & $11.70 \pm 0.15$\\
    $-107$    & $20$ & $15.00 \pm 0.04$ & $14.81 \pm 0.04$ & $12.65 \pm 0.09$ & $12.31 \pm 0.03$ & $14.21 \pm 0.01$ & $13.22 \pm 0.08$ & $12.64 \pm 0.02$\\
    $ -54$    & $10$ &      \ldots      & $14.00 \pm 0.19$ &      \ldots      &      \ldots      & $13.30 \pm 0.03$ &      \ldots      & $12.06 \pm 0.07$\\
    $   0$    & $22$ & $14.90 \pm 0.06$ & $15.02 \pm 0.02$ & $12.21 \pm 0.27$ & $12.50 \pm 0.02$ & $14.20 \pm 0.01$ & $13.10 \pm 0.16$ & $12.68 \pm 0.02$\\
    $  30$    & $19$ & $14.51 \pm 0.20$ &      \ldots      &      \ldots      &      \ldots      & $13.93 \pm 0.02$ & $13.03 \pm 0.29$ &      \ldots     \\
    $  54$    & $21$ & $14.71 \pm 0.11$ & $14.52 \pm 0.05$ & $12.67 \pm 0.10$ & $12.23 \pm 0.04$ & $14.12 \pm 0.01$ & $13.33 \pm 0.11$ & $12.43 \pm 0.03$\\
    $ 133$    & $17$ & $14.30 \pm 0.18$ & $14.14 \pm 0.10$ &      \ldots      & $11.50 \pm 0.20$ & $13.62 \pm 0.02$ & $12.81 \pm 0.17$ & $11.74 \pm 0.16$\\
    $ 161$    & $ 5$ &      \ldots      &      \ldots      &      \ldots      & $11.49 \pm 0.18$ & $13.39 \pm 0.03$ &      \ldots      & $11.48 \pm 0.24$\\
    $ 214$    & $35$ &      \ldots      &      \ldots      &      \ldots      &      \ldots      & $13.52 \pm 0.03$ &      \ldots      &      \ldots     \\
\hline
\end{tabular}
\end{center}

{\flushleft
$^{(a)}$ Relative to the systemic redshift $z_{\rm sys} = 2.13249$.\\}
\end{table*}

%% file: Krogager_Q2225_arXiv.bbl
\begin{thebibliography}{66}
\expandafter\ifx\csname natexlab\endcsname\relax\def\natexlab#1{#1}\fi

\bibitem[{{Arabsalmani} {et~al.}(2015){Arabsalmani}, {M{\o}ller}, {Fynbo},
  {Christensen}, {Freudling}, {Savaglio}, \& {Zafar}}]{Arabsalmani2015}
{Arabsalmani}, M., {M{\o}ller}, P., {Fynbo}, J.~P.~U., {Christensen}, L.,
  {Freudling}, W., {Savaglio}, S., \& {Zafar}, T., 2015, \mnras, 446, 990

\bibitem[{{Asplund} {et~al.}(2009){Asplund}, {Grevesse}, {Sauval}, \&
  {Scott}}]{Asplund2009}
{Asplund}, M., {Grevesse}, N., {Sauval}, A.~J., \& {Scott}, P., 2009, \araa,
  47, 481

\bibitem[{{Balashev} {et~al.}(2011){Balashev}, {Petitjean}, {Ivanchik},
  {Ledoux}, {Srianand}, {Noterdaeme}, \& {Varshalovich}}]{Balashev2011}
{Balashev}, S.~A., {Petitjean}, P., {Ivanchik}, A.~V., {Ledoux}, C.,
  {Srianand}, R., {Noterdaeme}, P., \& {Varshalovich}, D.~A., 2011, \mnras,
  418, 357

\bibitem[{{Balokovi{\'c}} {et~al.}(2012){Balokovi{\'c}}, {Smol{\v c}i{\'c}},
  {Ivezi{\'c}}, {Zamorani}, {Schinnerer}, \& {Kelly}}]{Balokovic2012}
{Balokovi{\'c}}, M., {Smol{\v c}i{\'c}}, V., {Ivezi{\'c}}, {\v Z}., {Zamorani},
  G., {Schinnerer}, E., \& {Kelly}, B.~C., 2012, \apj, 759, 30

\bibitem[{{Barthel} {et~al.}(1988){Barthel}, {Miley}, {Schilizzi}, \&
  {Lonsdale}}]{Barthel1988}
{Barthel}, P.~D., {Miley}, G.~K., {Schilizzi}, R.~T., \& {Lonsdale}, C.~J.,
  1988, \aaps, 73, 515

\bibitem[{{Barthel} {et~al.}(1990){Barthel}, {Tytler}, \&
  {Thomson}}]{Barthel1990}
{Barthel}, P.~D., {Tytler}, D.~R., \& {Thomson}, B., 1990, \aaps, 82, 339

\bibitem[{{Berg} {et~al.}(2015){Berg}, {Ellison}, {Prochaska}, {Venn}, \&
  {Dessauges-Zavadsky}}]{Berg2015}
{Berg}, T.~A.~M., {Ellison}, S.~L., {Prochaska}, J.~X., {Venn}, K.~A., \&
  {Dessauges-Zavadsky}, M., 2015, \mnras, 452, 4326

\bibitem[{{Christensen} {et~al.}(2014){Christensen}, {M{\o}ller}, {Fynbo}, \&
  {Zafar}}]{Christensen2014}
{Christensen}, L., {M{\o}ller}, P., {Fynbo}, J.~P.~U., \& {Zafar}, T., 2014,
  \mnras, 445, 225

\bibitem[{{Davis} {et~al.}(2007){Davis}, {Woo}, \& {Blaes}}]{Davis2007}
{Davis}, S.~W., {Woo}, J.-H., \& {Blaes}, O.~M., 2007, \apj, 668, 682

\bibitem[{{De Cia} {et~al.}(2013){De Cia}, {Ledoux}, {Savaglio}, {Schady}, \&
  {Vreeswijk}}]{DeCia2013}
{De Cia}, A., {Ledoux}, C., {Savaglio}, S., {Schady}, P., \& {Vreeswijk},
  P.~M., 2013, \aap, 560, A88

\bibitem[{{Ellison} {et~al.}(2004){Ellison}, {Churchill}, {Rix}, \&
  {Pettini}}]{Ellison2004}
{Ellison}, S.~L., {Churchill}, C.~W., {Rix}, S.~A., \& {Pettini}, M., 2004,
  \apj, 615, 118

\bibitem[{{Ellison} {et~al.}(2001){Ellison}, {Pettini}, {Steidel}, \&
  {Shapley}}]{Ellison2001a}
{Ellison}, S.~L., {Pettini}, M., {Steidel}, C.~C., \& {Shapley}, A.~E., 2001,
  \apj, 549, 770

\bibitem[{{Fynbo} {et~al.}(2013{\natexlab{a}}){Fynbo}, {Geier}, {Christensen},
  {Gallazzi}, {Krogager}, {Kr{\"u}hler}, {Ledoux}, {Maund}, {M{\o}ller},
  {Noterdaeme}, {Rivera-Thorsen}, \& {Vestergaard}}]{Fynbo2013b}
{Fynbo}, J.~P.~U., {Geier}, S.~J., {Christensen}, L., {et~al.},
  2013{\natexlab{a}}, \mnras, 436, 361

\bibitem[{{Fynbo} {et~al.}(2013{\natexlab{b}}){Fynbo}, {Krogager}, {Venemans},
  {Noterdaeme}, {Vestergaard}, {M{\o}ller}, {Ledoux}, \& {Geier}}]{Fynbo2013a}
{Fynbo}, J.~P.~U., {Krogager}, J.-K., {Venemans}, B., {Noterdaeme}, P.,
  {Vestergaard}, M., {M{\o}ller}, P., {Ledoux}, C., \& {Geier}, S.,
  2013{\natexlab{b}}, \apjs, 204, 6

\bibitem[{{Fynbo} {et~al.}(2011){Fynbo}, {Ledoux}, {Noterdaeme}, {Christensen},
  {M{\o}ller}, {Durgapal}, {Goldoni}, {Kaper}, {Krogager}, {Laursen}, {Maund},
  {Milvang-Jensen}, {Okoshi}, {Rasmussen}, {Thorsen}, {Toft}, \&
  {Zafar}}]{Fynbo2011}
{Fynbo}, J.~P.~U., {Ledoux}, C., {Noterdaeme}, P., {et~al.}, 2011, \mnras, 413,
  2481

\bibitem[{{Fynbo} {et~al.}(2008){Fynbo}, {Prochaska}, {Sommer-Larsen},
  {Dessauges-Zavadsky}, \& {M{\o}ller}}]{Fynbo2008}
{Fynbo}, J.~P.~U., {Prochaska}, J.~X., {Sommer-Larsen}, J.,
  {Dessauges-Zavadsky}, M., \& {M{\o}ller}, P., 2008, \apj, 683, 321

\bibitem[{{Gordon} {et~al.}(2003){Gordon}, {Clayton}, {Misselt}, {Landolt}, \&
  {Wolff}}]{Gordon03}
{Gordon}, K.~D., {Clayton}, G.~C., {Misselt}, K.~A., {Landolt}, A.~U., \&
  {Wolff}, M.~J., 2003, \apj, 594, 279

\bibitem[{{Gower} {et~al.}(1967){Gower}, {Scott}, \& {Wills}}]{Gower1967}
{Gower}, J.~F.~R., {Scott}, P.~F., \& {Wills}, D., 1967, \memras, 71, 49

\bibitem[{{Graham} {et~al.}(2014){Graham}, {Djorgovski}, {Drake}, {Mahabal},
  {Chang}, {Stern}, {Donalek}, \& {Glikman}}]{Graham2014}
{Graham}, M.~J., {Djorgovski}, S.~G., {Drake}, A.~J., {Mahabal}, A.~A.,
  {Chang}, M., {Stern}, D., {Donalek}, C., \& {Glikman}, E., 2014, \mnras, 439,
  703

\bibitem[{{Heckman} {et~al.}(1991){Heckman}, {Miley}, {Lehnert}, \& {van
  Breugel}}]{Heckman1991}
{Heckman}, T.~M., {Miley}, G.~K., {Lehnert}, M.~D., \& {van Breugel}, W., 1991,
  \apj, 370, 78

\bibitem[{{Heintz} {et~al.}(2015){Heintz}, {Fynbo}, \& {H{\o}g}}]{Heintz2015}
{Heintz}, K.~E., {Fynbo}, J.~P.~U., \& {H{\o}g}, E., 2015, \aap, 578, A91

\bibitem[{{Jorgenson} {et~al.}(2014){Jorgenson}, {Murphy}, {Thompson}, \&
  {Carswell}}]{Jorgenson2014}
{Jorgenson}, R.~A., {Murphy}, M.~T., {Thompson}, R., \& {Carswell}, R.~F.,
  2014, \mnras, 443, 2783

\bibitem[{{Jorgenson} {et~al.}(2006){Jorgenson}, {Wolfe}, {Prochaska}, {Lu},
  {Howk}, {Cooke}, {Gawiser}, \& {Gelino}}]{Jorgenson2006}
{Jorgenson}, R.~A., {Wolfe}, A.~M., {Prochaska}, J.~X., {Lu}, L., {Howk},
  J.~C., {Cooke}, J., {Gawiser}, E., \& {Gelino}, D.~M., 2006, \apj, 646, 730

\bibitem[{{Junkkarinen} {et~al.}(1991){Junkkarinen}, {Hewitt}, \&
  {Burbidge}}]{Junkkarinen1991}
{Junkkarinen}, V., {Hewitt}, A., \& {Burbidge}, G., 1991, \apjs, 77, 203

\bibitem[{{Kaspi} {et~al.}(2007){Kaspi}, {Brandt}, {Maoz}, {Netzer},
  {Schneider}, \& {Shemmer}}]{Kaspi2007}
{Kaspi}, S., {Brandt}, W.~N., {Maoz}, D., {Netzer}, H., {Schneider}, D.~P., \&
  {Shemmer}, O., 2007, \apj, 659, 997

\bibitem[{{Kennicutt}(1998)}]{Kennicutt1998}
{Kennicutt}, Jr., R.~C., 1998, \araa, 36, 189

\bibitem[{{Khare} {et~al.}(2012){Khare}, {vanden Berk}, {York}, {Lundgren}, \&
  {Kulkarni}}]{Khare2012}
{Khare}, P., {vanden Berk}, D., {York}, D.~G., {Lundgren}, B., \& {Kulkarni},
  V.~P., 2012, \mnras, 419, 1028

\bibitem[{{Krawczyk} {et~al.}(2015){Krawczyk}, {Richards}, {Gallagher},
  {Leighly}, {Hewett}, {Ross}, \& {Hall}}]{Krawczyk2015}
{Krawczyk}, C.~M., {Richards}, G.~T., {Gallagher}, S.~C., {Leighly}, K.~M.,
  {Hewett}, P.~C., {Ross}, N.~P., \& {Hall}, P.~B., 2015, \aj, 149, 203

\bibitem[{{Krogager} {et~al.}(2013){Krogager}, {Fynbo}, {Ledoux},
  {Christensen}, {Gallazzi}, {Laursen}, {M{\o}ller}, {Noterdaeme}, {Peroux},
  {Pettini}, \& {Vestergaard}}]{Krogager2013}
{Krogager}, J.-K., {Fynbo}, J.~P.~U., {Ledoux}, C., {et~al.}, 2013, \mnras,
  433, 3091

\bibitem[{{Krogager} {et~al.}(2012){Krogager}, {Fynbo}, {M{\o}ller}, {Ledoux},
  {Noterdaeme}, {Christensen}, {Milvang-Jensen}, \& {Sparre}}]{Krogager2012}
{Krogager}, J.-K., {Fynbo}, J.~P.~U., {M{\o}ller}, P., {Ledoux}, C.,
  {Noterdaeme}, P., {Christensen}, L., {Milvang-Jensen}, B., \& {Sparre}, M.,
  2012, \mnras, 424, L1

\bibitem[{{Krogager} {et~al.}(2015){Krogager}, {Geier}, {Fynbo}, {Venemans},
  {Ledoux}, {M{\o}ller}, {Noterdaeme}, {Vestergaard}, {Kangas}, {Pursimo},
  {Saturni}, \& {Smirnova}}]{Krogager2015}
{Krogager}, J.-K., {Geier}, S., {Fynbo}, J.~P.~U., {et~al.}, 2015, \apjs, 217,
  5

\bibitem[{{Ledoux} {et~al.}(2015){Ledoux}, {Noterdaeme}, {Petitjean}, \&
  {Srianand}}]{Ledoux2015}
{Ledoux}, C., {Noterdaeme}, P., {Petitjean}, P., \& {Srianand}, R., 2015, \aap,
  580, A8

\bibitem[{{Ledoux} {et~al.}(2006){Ledoux}, {Petitjean}, {Fynbo}, {M{\o}ller},
  \& {Srianand}}]{Ledoux2006}
{Ledoux}, C., {Petitjean}, P., {Fynbo}, J.~P.~U., {M{\o}ller}, P., \&
  {Srianand}, R., 2006, \aap, 457, 71

\bibitem[{{Liszt}(2014)}]{Liszt2014}
{Liszt}, H., 2014, \apj, 783, 17

\bibitem[{{M{\o}ller} {et~al.}(2004){M{\o}ller}, {Fynbo}, \&
  {Fall}}]{Moller2004}
{M{\o}ller}, P., {Fynbo}, J.~P.~U., \& {Fall}, S.~M., 2004, \aap, 422, L33

\bibitem[{{M{\o}ller} {et~al.}(2013){M{\o}ller}, {Fynbo}, {Ledoux}, \&
  {Nilsson}}]{Moller2013}
{M{\o}ller}, P., {Fynbo}, J.~P.~U., {Ledoux}, C., \& {Nilsson}, K.~K., 2013,
  \mnras, 430, 2680

\bibitem[{{Neeleman} {et~al.}(2013){Neeleman}, {Wolfe}, {Prochaska}, \&
  {Rafelski}}]{Neeleman2013}
{Neeleman}, M., {Wolfe}, A.~M., {Prochaska}, J.~X., \& {Rafelski}, M., 2013,
  \apj, 769, 54

\bibitem[{{Noterdaeme} {et~al.}(2012){Noterdaeme}, {Laursen}, {Petitjean},
  {Vergani}, {Maureira}, {Ledoux}, {Fynbo}, {L{\'o}pez}, \&
  {Srianand}}]{Noterdaeme2012a}
{Noterdaeme}, P., {Laursen}, P., {Petitjean}, P., {et~al.}, 2012, \aap, 540,
  A63

\bibitem[{{Noterdaeme} {et~al.}(2008){Noterdaeme}, {Ledoux}, {Petitjean}, \&
  {Srianand}}]{Noterdaeme2008}
{Noterdaeme}, P., {Ledoux}, C., {Petitjean}, P., \& {Srianand}, R., 2008, \aap,
  481, 327

\bibitem[{{Noterdaeme} {et~al.}(2010){Noterdaeme}, {Petitjean}, {Ledoux},
  {L{\'o}pez}, {Srianand}, \& {Vergani}}]{Noterdaeme2010}
{Noterdaeme}, P., {Petitjean}, P., {Ledoux}, C., {L{\'o}pez}, S., {Srianand},
  R., \& {Vergani}, S.~D., 2010, \aap, 523, A80+

\bibitem[{{Noterdaeme} {et~al.}(2015){Noterdaeme}, {Petitjean}, \&
  {Srianand}}]{Noterdaeme2015b}
{Noterdaeme}, P., {Petitjean}, P., \& {Srianand}, R., 2015, \aap, 578, L5

\bibitem[{{Pei} {et~al.}(1991){Pei}, {Fall}, \& {Bechtold}}]{Pei1991}
{Pei}, Y.~C., {Fall}, S.~M., \& {Bechtold}, J., 1991, \apj, 378, 6

\bibitem[{{Pei} {et~al.}(1999){Pei}, {Fall}, \& {Hauser}}]{Pei1999}
{Pei}, Y.~C., {Fall}, S.~M., \& {Hauser}, M.~G., 1999, \apj, 522, 604

\bibitem[{{Peth} {et~al.}(2011){Peth}, {Ross}, \& {Schneider}}]{Peth2011}
{Peth}, M.~A., {Ross}, N.~P., \& {Schneider}, D.~P., 2011, \aj, 141, 105

\bibitem[{{Planck Collaboration} {et~al.}(2014){Planck Collaboration}, {Ade},
  {Aghanim}, {Armitage-Caplan}, {Arnaud}, {Ashdown}, {Atrio-Barandela},
  {Aumont}, {Baccigalupi}, {Banday}, \& et~al.}]{Planck2014}
{Planck Collaboration}, {Ade}, P.~A.~R., {Aghanim}, N., {et~al.}, 2014, \aap,
  571, A16

\bibitem[{{Pontzen} \& {Pettini}(2009)}]{Pontzen2009}
{Pontzen}, A. \& {Pettini}, M., 2009, \mnras, 393, 557

\bibitem[{{Prochaska} \& {Wolfe}(1997)}]{Prochaska1997}
{Prochaska}, J.~X. \& {Wolfe}, A.~M., 1997, \apj, 487, 73

\bibitem[{{Rafelski} {et~al.}(2012){Rafelski}, {Wolfe}, {Prochaska},
  {Neeleman}, \& {Mendez}}]{Rafelski2012}
{Rafelski}, M., {Wolfe}, A.~M., {Prochaska}, J.~X., {Neeleman}, M., \&
  {Mendez}, A.~J., 2012, \apj, 755, 89

\bibitem[{{Richards} {et~al.}(2002){Richards}, {Fan}, {Newberg}, {Strauss},
  {Vanden Berk}, {Schneider}, {Yanny}, {Boucher}, {Burles}, {Frieman}, {Gunn},
  {Hall}, {Ivezi{\'c}}, {Kent}, {Loveday}, {Lupton}, {Rockosi}, {Schlegel},
  {Stoughton}, {SubbaRao}, \& {York}}]{Richards2002}
{Richards}, G.~T., {Fan}, X., {Newberg}, H.~J., {et~al.}, 2002, \aj, 123, 2945

\bibitem[{{Richards} {et~al.}(2001){Richards}, {Fan}, {Schneider}, {Vanden
  Berk}, {Strauss}, {York}, {Anderson}, {Anderson}, {Annis}, {Bahcall},
  {Bernardi}, {Briggs}, {Brinkmann}, {Brunner}, {Burles}, {Carey}, {Castander},
  {Connolly}, {Crocker}, {Csabai}, {Doi}, {Finkbeiner}, {Friedman}, {Frieman},
  {Fukugita}, {Gunn}, {Hindsley}, {Ivezi{\'c}}, {Kent}, {Knapp}, {Lamb},
  {Leger}, {Long}, {Loveday}, {Lupton}, {McKay}, {Meiksin}, {Merrelli}, {Munn},
  {Newberg}, {Newcomb}, {Nichol}, {Owen}, {Pier}, {Pope}, {Richmond},
  {Rockosi}, {Schlegel}, {Siegmund}, {Smee}, {Snir}, {Stoughton}, {Stubbs},
  {SubbaRao}, {Szalay}, {Szokoly}, {Tremonti}, {Uomoto}, {Waddell}, {Yanny}, \&
  {Zheng}}]{Richards2001}
{Richards}, G.~T., {Fan}, X., {Schneider}, D.~P., {et~al.}, 2001, \aj, 121,
  2308

\bibitem[{{Ross} {et~al.}(2012){Ross}, {Myers}, {Sheldon}, {Y{\`e}che},
  {Strauss}, {Bovy}, {Kirkpatrick}, {Richards}, {Aubourg}, {Blanton}, {Brandt},
  {Carithers}, {Croft}, {da Silva}, {Dawson}, {Eisenstein}, {Hennawi}, {Ho},
  {Hogg}, {Lee}, {Lundgren}, {McMahon}, {Miralda-Escud{\'e}},
  {Palanque-Delabrouille}, {P{\^a}ris}, {Petitjean}, {Pieri}, {Rich}, {Roe},
  {Schiminovich}, {Schlegel}, {Schneider}, {Slosar}, {Suzuki}, {Tinker},
  {Weinberg}, {Weyant}, {White}, \& {Wood-Vasey}}]{Ross2012}
{Ross}, N.~P., {Myers}, A.~D., {Sheldon}, E.~S., {et~al.}, 2012, \apjs, 199, 3

\bibitem[{{Ryabinkov} {et~al.}(2003){Ryabinkov}, {Kaminker}, \&
  {Varshalovich}}]{Ryabinkov2003}
{Ryabinkov}, A.~I., {Kaminker}, A.~D., \& {Varshalovich}, D.~A., 2003, \aap,
  412, 707

\bibitem[{{Schlafly} \& {Finkbeiner}(2011)}]{Schlafly2011}
{Schlafly}, E.~F. \& {Finkbeiner}, D.~P., 2011, \apj, 737, 103

\bibitem[{{Schmidt} {et~al.}(2010){Schmidt}, {Marshall}, {Rix}, {Jester},
  {Hennawi}, \& {Dobler}}]{Schmidt2010}
{Schmidt}, K.~B., {Marshall}, P.~J., {Rix}, H.-W., {Jester}, S., {Hennawi},
  J.~F., \& {Dobler}, G., 2010, \apj, 714, 1194

\bibitem[{{Selsing} {et~al.}(2015){Selsing}, {Fynbo}, {Christensen}, \&
  {Krogager}}]{Selsing2015_prep}
{Selsing}, J., {Fynbo}, J. P.~U., {Christensen}, L., \& {Krogager}, J.-K.,
  2015, \aap, submitted

\bibitem[{{Srianand} {et~al.}(2008){Srianand}, {Noterdaeme}, {Ledoux}, \&
  {Petitjean}}]{Srianand2008a}
{Srianand}, R., {Noterdaeme}, P., {Ledoux}, C., \& {Petitjean}, P., 2008, \aap,
  482, L39

\bibitem[{{Vanden Berk} {et~al.}(2001){Vanden Berk}, {Richards}, {Bauer},
  {Strauss}, {Schneider}, {Heckman}, {York}, {Hall}, {Fan}, {Knapp},
  {Anderson}, {Annis}, {Bahcall}, {Bernardi}, {Briggs}, {Brinkmann}, {Brunner},
  {Burles}, {Carey}, {Castander}, {Connolly}, {Crocker}, {Csabai}, {Doi},
  {Finkbeiner}, {Friedman}, {Frieman}, {Fukugita}, {Gunn}, {Hennessy},
  {Ivezi{\'c}}, {Kent}, {Kunszt}, {Lamb}, {Leger}, {Long}, {Loveday}, {Lupton},
  {Meiksin}, {Merelli}, {Munn}, {Newberg}, {Newcomb}, {Nichol}, {Owen}, {Pier},
  {Pope}, {Rockosi}, {Schlegel}, {Siegmund}, {Smee}, {Snir}, {Stoughton},
  {Stubbs}, {SubbaRao}, {Szalay}, {Szokoly}, {Tremonti}, {Uomoto}, {Waddell},
  {Yanny}, \& {Zheng}}]{vandenBerk2001}
{Vanden Berk}, D.~E., {Richards}, G.~T., {Bauer}, A., {et~al.}, 2001, \aj, 122,
  549

\bibitem[{{Vernet} {et~al.}(2011){Vernet}, {Dekker}, {D'Odorico}, {Kaper},
  {Kjaergaard}, {Hammer}, {Randich}, {Zerbi}, {Groot}, {Hjorth}, {Guinouard},
  {Navarro}, {Adolfse}, {Albers}, {Amans}, {Andersen}, {Andersen}, {Binetruy},
  {Bristow}, {Castillo}, {Chemla}, {Christensen}, {Conconi}, {Conzelmann},
  {Dam}, {de Caprio}, {de Ugarte Postigo}, {Delabre}, {di Marcantonio},
  {Downing}, {Elswijk}, {Finger}, {Fischer}, {Flores}, {Fran{\c c}ois},
  {Goldoni}, {Guglielmi}, {Haigron}, {Hanenburg}, {Hendriks}, {Horrobin},
  {Horville}, {Jessen}, {Kerber}, {Kern}, {Kiekebusch}, {Kleszcz}, {Klougart},
  {Kragt}, {Larsen}, {Lizon}, {Lucuix}, {Mainieri}, {Manuputy}, {Martayan},
  {Mason}, {Mazzoleni}, {Michaelsen}, {Modigliani}, {Moehler}, {M{\o}ller},
  {Norup S{\o}rensen}, {N{\o}rregaard}, {P{\'e}roux}, {Patat}, {Pena}, {Pragt},
  {Reinero}, {Rigal}, {Riva}, {Roelfsema}, {Royer}, {Sacco}, {Santin},
  {Schoenmaker}, {Spano}, {Sweers}, {Ter Horst}, {Tintori}, {Tromp}, {van
  Dael}, {van der Vliet}, {Venema}, {Vidali}, {Vinther}, {Vola}, {Winters},
  {Wistisen}, {Wulterkens}, \& {Zacchei}}]{Vernet2011}
{Vernet}, J., {Dekker}, H., {D'Odorico}, S., {et~al.}, 2011, \aap, 536, A105

\bibitem[{{Vestergaard} \& {Wilkes}(2001)}]{Vestergaard2001}
{Vestergaard}, M. \& {Wilkes}, B.~J., 2001, \apjs, 134, 1

\bibitem[{{Vladilo} {et~al.}(2006){Vladilo}, {Centuri{\'o}n}, {Levshakov},
  {P{\'e}roux}, {Khare}, {Kulkarni}, \& {York}}]{Vladilo2006}
{Vladilo}, G., {Centuri{\'o}n}, M., {Levshakov}, S.~A., {P{\'e}roux}, C.,
  {Khare}, P., {Kulkarni}, V.~P., \& {York}, D.~G., 2006, \aap, 454, 151

\bibitem[{{Vladilo} \& {P{\'e}roux}(2005)}]{Vladilo2005}
{Vladilo}, G. \& {P{\'e}roux}, C., 2005, \aap, 444, 461

\bibitem[{{Vladilo} {et~al.}(2008){Vladilo}, {Prochaska}, \&
  {Wolfe}}]{Vladilo2008}
{Vladilo}, G., {Prochaska}, J.~X., \& {Wolfe}, A.~M., 2008, \aap, 478, 701

\bibitem[{{Wang} {et~al.}(2012){Wang}, {Zhou}, {Ge}, {Jiang}, {Lu},
  {Prochaska}, {Hamann}, {Wang}, {Wang}, \& {Yuan}}]{Wang2012}
{Wang}, J.-G., {Zhou}, H.-Y., {Ge}, J., {et~al.}, 2012, \apj, 760, 42

\bibitem[{{Wolfe} {et~al.}(2005){Wolfe}, {Gawiser}, \& {Prochaska}}]{Wolfe2005}
{Wolfe}, A.~M., {Gawiser}, E., \& {Prochaska}, J.~X., 2005, \araa, 43, 861

\bibitem[{{York} {et~al.}(2000){York}, {Adelman}, {Anderson}, {Anderson},
  {Annis}, {Bahcall}, {Bakken}, {Barkhouser}, {Bastian}, {Berman}, {Boroski},
  {Bracker}, {Briegel}, {Briggs}, {Brinkmann}, {Brunner}, {Burles}, {Carey},
  {Carr}, {Castander}, {Chen}, {Colestock}, {Connolly}, {Crocker}, {Csabai},
  {Czarapata}, {Davis}, {Doi}, {Dombeck}, {Eisenstein}, {Ellman}, {Elms},
  {Evans}, {Fan}, {Federwitz}, {Fiscelli}, {Friedman}, {Frieman}, {Fukugita},
  {Gillespie}, {Gunn}, {Gurbani}, {de Haas}, {Haldeman}, {Harris}, {Hayes},
  {Heckman}, {Hennessy}, {Hindsley}, {Holm}, {Holmgren}, {Huang}, {Hull},
  {Husby}, {Ichikawa}, {Ichikawa}, {Ivezi{\'c}}, {Kent}, {Kim}, {Kinney},
  {Klaene}, {Kleinman}, {Kleinman}, {Knapp}, {Korienek}, {Kron}, {Kunszt},
  {Lamb}, {Lee}, {Leger}, {Limmongkol}, {Lindenmeyer}, {Long}, {Loomis},
  {Loveday}, {Lucinio}, {Lupton}, {MacKinnon}, {Mannery}, {Mantsch}, {Margon},
  {McGehee}, {McKay}, {Meiksin}, {Merelli}, {Monet}, {Munn}, {Narayanan},
  {Nash}, {Neilsen}, {Neswold}, {Newberg}, {Nichol}, {Nicinski}, {Nonino},
  {Okada}, {Okamura}, {Ostriker}, {Owen}, {Pauls}, {Peoples}, {Peterson},
  {Petravick}, {Pier}, {Pope}, {Pordes}, {Prosapio}, {Rechenmacher}, {Quinn},
  {Richards}, {Richmond}, {Rivetta}, {Rockosi}, {Ruthmansdorfer}, {Sandford},
  {Schlegel}, {Schneider}, {Sekiguchi}, {Sergey}, {Shimasaku}, {Siegmund},
  {Smee}, {Smith}, {Snedden}, {Stone}, {Stoughton}, {Strauss}, {Stubbs},
  {SubbaRao}, {Szalay}, {Szapudi}, {Szokoly}, {Thakar}, {Tremonti}, {Tucker},
  {Uomoto}, {Vanden Berk}, {Vogeley}, {Waddell}, {Wang}, {Watanabe},
  {Weinberg}, {Yanny}, {Yasuda}, \& {SDSS Collaboration}}]{York2000}
{York}, D.~G., {Adelman}, J., {Anderson}, Jr., J.~E., {et~al.}, 2000, \aj, 120,
  1579

\bibitem[{{Zafar} {et~al.}(2015){Zafar}, {M{\o}ller}, {Watson}, {Fynbo},
  {Krogager}, {Zafar}, {Saturni}, {Geier}, \& {Venemans}}]{Zafar2015_prep}
{Zafar}, T., {M{\o}ller}, P., {Watson}, D., {et~al.}, 2015, \aap, accepted,
  arXiv:1510.01708

\end{thebibliography}
